\documentclass[twocolumn]{aastex63}
\usepackage{bm}

\accepted{April 26, 2021}
\submitjournal{ApJ}
\shorttitle{The environment of multiple galaxy populations}
\shortauthors{Ito et al.}
\graphicspath{{./}{figures/}}

\begin{document}

\title{Interrelation of the environment of Ly$\alpha$ emitters and massive galaxies at $2<z<4.5$}
\correspondingauthor{Kei Ito}
\email{kei.ito@grad.nao.ac.jp}
\author[0000-0002-9453-0381]{Kei Ito}
\affiliation{Department of Astronomical Science, The Graduate University for Advanced Studies, SOKENDAI, Mitaka, Tokyo, 181-8588, Japan} 
\affiliation{National Astronomical Observatory of Japan, Mitaka, Tokyo, 181-8588, Japan}
\affiliation{Department of Astronomy, School of Science, The University of Tokyo, 7-3-1 Hongo, Bunkyo-ku, Tokyo, 113-0033, Japan}
\author[0000-0003-3954-4219]{Nobunari Kashikawa}
\affiliation{Department of Astronomy, School of Science, The University of Tokyo, 7-3-1 Hongo, Bunkyo-ku, Tokyo, 113-0033, Japan}
\affiliation{Research Center for the Early Universe, The University of Tokyo, 7-3-1 Hongo, Bunkyo-ku, Tokyo 113-0033, Japan}
\author{Masayuki Tanaka}
\affiliation{National Astronomical Observatory of Japan, Mitaka, Tokyo, 181-8588, Japan}
\affiliation{Department of Astronomical Science, The Graduate University for Advanced Studies, SOKENDAI, Mitaka, Tokyo, 181-8588, Japan}
\author{Mariko Kubo}
\affiliation{Research Center for Space and Cosmic Evolution, Ehime University, Matsuyama, Ehime 790-8577, Japan}
\author[0000-0002-2725-302X]{Yongming Liang}
\affiliation{Department of Astronomical Science, The Graduate University for Advanced Studies, SOKENDAI, Mitaka, Tokyo, 181-8588, Japan} 
\affiliation{National Astronomical Observatory of Japan, Mitaka, Tokyo, 181-8588, Japan}
\affiliation{Department of Astronomy, School of Science, The University of Tokyo, 7-3-1 Hongo, Bunkyo-ku, Tokyo, 113-0033, Japan}
\author{Jun Toshikawa}
\affiliation{Institute for Cosmic Ray Research, The University of Tokyo, 5-1-5 Kashiwa-no-Ha, Kashiwa, Chiba, 277-8582, Japan}
\affiliation{Department of Physics, University of Bath, Claverton Down, Bath, BA2 7AY, UK}
\author{Hisakazu Uchiyama}
\affiliation{National Astronomical Observatory of Japan, Mitaka, Tokyo, 181-8588, Japan}
\author{Rikako Ishimoto}
\affiliation{Department of Astronomy, School of Science, The University of Tokyo, 7-3-1 Hongo, Bunkyo-ku, Tokyo, 113-0033, Japan}
\author{Takehiro Yoshioka}
\affiliation{Department of Astronomy, School of Science, The University of Tokyo, 7-3-1 Hongo, Bunkyo-ku, Tokyo, 113-0033, Japan}
\author{Yoshihiro Takeda}
\affiliation{Department of Astronomy, School of Science, The University of Tokyo, 7-3-1 Hongo, Bunkyo-ku, Tokyo, 113-0033, Japan}

\begin{abstract}
We present a comparison of the spatial distributions of Ly$\alpha$ emitters (LAEs) and massive star-forming and quiescent galaxies (SFGs and QGs) at $2<z<4.5$. We use the photometric redshift catalog to select SFGs and QGs and a LAE catalog from intermediate/narrow bands obtained from the Subaru Telescope and Isaac-Newton Telescope in Cosmic Evolution Survey (COSMOS). We derive the auto-/cross- correlation signals of SFGs, QGs, and LAEs, and the galaxy overdensity distributions at the position of them. Whereas the cross-correlation signals of SFGs and QGs are explained solely by their halo mass differences, those of SFGs and LAEs are significantly lower than those expected from their auto-correlation signals, suggesting that some additional physical processes are segregating these two populations. Such segregation of SFGs and LAEs becomes stronger for rest-frame ultraviolet faint LAEs ($M_{\rm UV}>-20$). From the overdensity distributions, LAEs are located in less dense regions than SFGs and QGs, whereas SFGs and QGs tend to be in the same overdensity distributions. The different spatial distributions of LAEs compared to those of massive galaxies may be attributed to assembly bias or large amounts of neutral hydrogen gas associated with massive halos. These results reinforce the importance of exploring multiple galaxy populations in quantifying the intrinsic galaxy environment of the high-$z$ universe.
\end{abstract}

\section{Introduction} \label{sec:1}
\par In the last few decades, galaxies at $z\geq2$ have been energetically explored. To select high redshift (high-$z$) galaxies, a large number of studies use continuum emission. The photometric redshift (photo-$z$) and the spectral energy distribution (SED) of multiple photometric data can classify star-forming galaxies (SFGs) and quiescent galaxies (QGs) \citep[e.g.,][]{Ilbert2009,Laigle2016,Mortlock2015,Mawatari2020}. In particular, QGs have been selected photometrically up to $z\sim6$ \citep[e.g.,][]{Ilbert2013,Davidzon2017,Merlin2019}, and recent near-infrared spectroscopy observations confirm their existence up to $z\sim4$ \citep[e.g.,][]{Glazebrook2017,Tanaka2019,Belli2019,Forrest2020,Valentino2020}.
\par Another powerful tool to explore the high-$z$ universe is line emitting galaxies. Ly$\alpha$ emitters (LAEs) are one of the most famous populations. LAEs are often selected through the excess in the narrow or intermediate band and are known to have lower stellar masses ($\log{(M_\star/M_\odot)}\sim9$) \citep[e.g.,][]{Santos2020}, younger stellar populations \citep[e.g.,][]{Gawiser2007,Hagen2014}, and less dust obscuration \citep[e.g.,][]{Kusakabe2015,Gawiser2006} than continuum selected galaxies. In terms of their host dark matter halos, LAEs tend to reside in halos with lower mass than more massive galaxies \citep[e.g.,][]{Khostovan2019,Kusakabe2018,Ouchi2018}. Another important population of line emitting galaxies are H$\alpha$ emitters (HAEs) \citep[e.g.,][]{Geach2008,Sobral2013,Kodama2013}. Unlike LAEs, they are known to be SFGs with similar properties as continuum selected ones \citep[e.g.,][]{Oteo2015}.
\par One of the essential topics to be investigated in the high-$z$ universe is connections between galaxy properties and their surrounding environments, referred to as the ``environmental effect". At lower redshifts ($z<1$), many studies report significant correlations \citep[e.g.,][]{Dressler1980,Peng2010}. To understand its origin, exploration in the higher-$z$ universe is vital. Indeed, recent statistical studies have discussed the environmental effect at $z>2$. In particular, the relation between the local density and the star formation rate (SFR) \citep[e.g.,][]{Lemaux2020,Chartab2020} and the environmental quenching \citep[e.g.,][]{Lin2016} have been intensively investigated, though no consensus has been obtained. \citet{Lemaux2020} suggest a positive correlation between SFR and the overdensity, whereas \citet{Chartab2020} report a negative correlation. Also, \citet{Lin2016} do not find any strong dependence of the environment for the quiescent fraction at $z>1.5$. 
\par The characterization of the environment may depend on the population of tracer galaxies. The connection between galaxies properties and the large scale environment at high-$z$ has also been investigated by focusing on progenitors of galaxy clusters in the local universe, referred to as ''protoclusters" \citep[e.g.,][]{Hatch2011,Koyama2013,Cooke2014,Ito2020,Shimakawa2018,Tadaki2019}. Protoclusters are often discovered through the projected number density of galaxies \citep[e.g.,][]{Steidel1998, Hatch2011, Jiang2018,Toshikawa2018}, but most protoclusters are selected using only SFGs. Whereas some studies report the high number density of multiple galaxy populations in protoclusters \citep[e.g.,][]{Kubo2013}, some studies have reported that LAEs and massive SFGs trace different large-scale structures \citep[e.g.,][]{Shimakawa2017,Shi2019,Shi2020}. Such possible segregation of LAEs and massive galaxies can be related to the assembly time difference of host dark matter halo \citep[e.g.,][]{Shi2019} or baryonic physics, such as the relation of surrounding H{\sc i} gas and Ly$\alpha$ emission \citep[e.g.,][]{Shimakawa2017}. These observational results suggest the importance of comprehensively understanding the interrelationship of the distributions of multiple populations.
\par For quantifying the differences of the spatial distribution of multiple populations, the cross-correlation is an effective tool. It has been measured for different galaxy populations \citep[e.g.,][]{Bethermin2014,Hatfield2017} and used to determine the connections between galaxies and intergalactic media (IGM) \citep[e.g.,][]{Tejos2014,Momose2020,Liang2020}. The overdensity, which is defined as the excess of the surface number density over the average value, is another tool for quantifying the environment. The overdensity is measured using a variety of methods. Some methods fix the scale in which the density is estimated, whereas others calculate the density based on the number of nearby galaxies. Recent studies have proposed a density measurement technique, called the Voronoi Monte Carlo Mapping, that does not assume any density scale \citep[e.g.,][]{Tomczak2017, Lemaux2020}. These methods are based on different assumptions, so it is essential to systematically apply a unified method to all populations to see the difference of their environments.
\par In this study, we examine the spatial distribution differences among multiple galaxy populations via analysis of their clustering and overdensity distributions to understand the general relationships among them in the large-scale structure. We focus on three populations: massive SFGs, massive QGs, and LAEs in the Cosmic Evolution Survey (COSMOS) field. SFGs and QGs are selected based on the multi-photometry catalog constructed in \citet{Laigle2016}, whereas LAEs are taken from the extensive narrow- and medium-band based survey in SC4K \citep{Sobral2018}. The large amounts of data in the COSMOS field enables us to investigate the differences in their distributions up to $z\sim4.5$.
\par This paper consists of the following sections. In Section 2, we introduce our galaxy samples. In Section 3, we present our clustering analysis. In Section 4, we report a comparison of the overdensity distribution at the positions of sources in the different populations. In Section 5, we examine the result by using HAEs at $z=2.22$ and discuss the implications of our results to the environments of galaxies. Lastly, in Section 6, we summarize our paper. 
\par In this study, we assume that the cosmological parameters are $H_0 = 70\ {\rm km\ s^{-1}\ Mpc^{-1}}= 100h\ {\rm km\ s^{-1}\ Mpc^{-1}}$, $\Omega_m$ = 0.3, and $\Omega_\Lambda = 0.7$, and use the AB magnitude system. We also use the annotations cMpc and pMpc to refer comoving and physical scales, respectively.
\section{Dataset}
\subsection{Star-forming galaxies and Quiescent galaxies}
\subsubsection{Sample construction}
\par We use the COSMOS multi-band catalog constructed in \citet{Laigle2016}. This catalog consists of photometries of $\sim 30$ bands, i.e., near-UV of GALEX \citep{Zamojski2007}, $u^*$-band of Canada--France--Hawaii Telescope, $BVri^+z^{++}$ and several intermediate/narrow bands of Suprime-Cam \citep{Taniguchi2007,Taniguchi2015} and Y-band of Hyper Suprime-Cam of Subaru Telescope, $YJHK_s$ of VISTA InfraRed Camera (VIRCAM) of the VISTA telescope \citep{McCracken2012}, $HK_s$ of Wide-field InfraRed  Camera (WIRCam) from Canada--France--Hawaii Telescope \citep{McCracken2009}, and Channels 1, 2, 3, and 4 of the Infrared Array Camera (IRAC) of the Spitzer telescope from the SPLASH survey. This catalog is based on detection in the $\chi^2$ sum of the $YJHK_s$ and $z^{++}$ images. For more details, please refer to \citet{Laigle2016}.
\par In this study, SFGs and QGs samples are constructed to be magnitude-limited based on the $3\sigma$ limiting magnitude of $K_s$ band ($K_s<24$), where $K_s$ is the $3\arcsec$ aperture magnitude of $K_s$ band. Objects are further selected with flags ({\tt FLAG\_COSMOS} and {\tt FLAG\_PETER}) to only focus on objects with the clean photometry. We then estimate the photometric redshift using {\tt MIZUKI} code \citep{Tanaka2015}. One advantage of this code is that we are able to simultaneously derive the photometric redshifts and physical properties (e.g., $M_\star$, SFR, and dust extinction) with their Bayesian priors. This allows us to include the uncertainty of the photometric redshift in the estimate of the physical properties. It should be noted that the photo-$z$ between {\tt MIZUKI} and \citet{Laigle2016} is consistent with each other with $\delta z/(1+z)=0.003$, after excluding objects with the bad chi-squares ($\chi^2_\nu>4$) at $2<z<4.5$. 
\par {\tt MIZUKI} conducts the fitting based on spectral templates from \citet{Bruzual2003}, Chabrier IMF \citep{Chabrier2003}, and Calzetti dust attenuation curve \citep{Calzetti2000}. We use an exponentially declining SFR, i.e., ${\rm SFR} \propto \exp(-t/\tau)$, where $t$ is time. The $\tau$ is assumed to be $0.1\ {\rm Gyr}<\tau<11\ {\rm Gyr}$ in addition to $\tau=0,\ \infty$, which is equivalent to the single stellar population model and the constant SFR model. The age is assumed to be between 0.05 and 14 Gyr. Also, the optical depth in the V band ($\tau_{V}$) is between 0 and 2 with a step of 0.1, in addition to $\tau_V=2.5,\ 3,\ {\rm and}\ 4$. Because the templates mentioned above include only stellar emissions, the nebular emissions are included according to \citet{Inoue2011}. Similar to \citet{Kubo2018}, which select QGs at $3.5<z<4.5$ in the ultradeep survey (UDS) region using {\tt MIZUKI} code, galaxies with bad chi-squares ($\chi^2_\nu>4$) in the SED fitting are excluded ($\sim4$\% of the total sample in the target redshift). Objects with large reduced chi-squares have generally poor-photometry due to the affection by nearby bright stars or blending. Also, apparent AGNs can be excluded from this criteria since any AGN template is not included in the fitting. The typical uncertainty of the estimated redshift, $M_\star$, and SFR of objects at $2<z<4.5$ are $\delta z/(1+z) \sim0.05$, $\delta M_\star/M_\star \sim0.2$, and $\delta {\rm SFR}/{\rm SFR}\sim 0.4$, respectively.
\par SFGs and QGs are distinguished based on the specific star formation rate (sSFR) derived by the SED fitting, as in our other studies \citep[][Ito et al. in prep.]{Kubo2018}. We define galaxies with ${\rm sSFR_{1\sigma, upper}}<10^{-9.5}\ {\rm yr^{-1}}$ as QGs, and classify the others as SFGs. Here, ${\rm sSFR_{1\sigma, upper}}$ is the upper limit of sSFR, which is defined as the ratio of the $1\sigma$ upper limit of SFR to the $1\sigma$ lower limit of stellar mass, derived from the SED fitting. We note that our result does not change even if we modify this classification, for example, by considering the redshift evolution of the star formation main sequence (i.e., a stricter threshold for lower-$z$ objects) or by defining SFGs with ${\rm sSFR_{1\sigma, lower}}>10^{-9.5}\ {\rm yr^{-1}}$ to exclude overlapped region with QGs.
\par
\begin{figure}
    \centering
    \includegraphics[width=8.5cm]{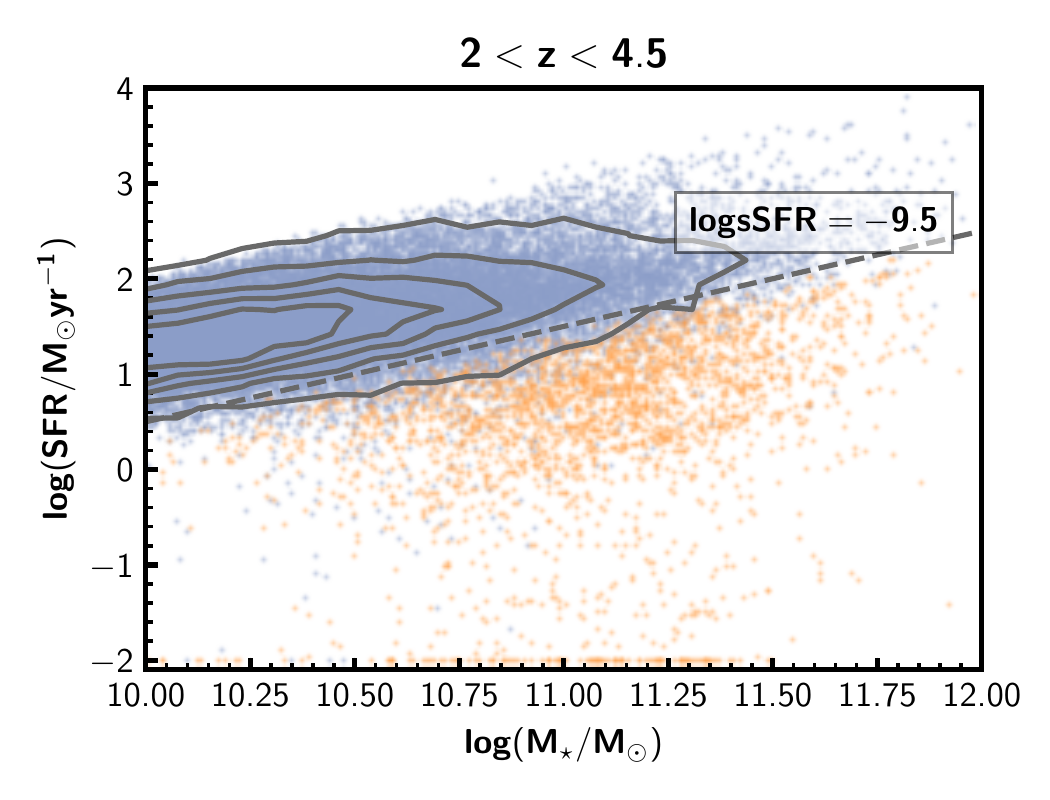}
    \caption{$M_\star$--SFR relation of target galaxies. The blue dots show SFGs, and the orange dots show QGs. Objects with SFR lower than $0.01M_\sun{\rm yr^{-1}}$ are shown at $0.01M_\sun{\rm yr^{-1}}$ for the illustrative purposes. For reference, the contour estimated from all objects in figure is shown. The dashed line shows the location of $\log{\rm (sSFR/{\rm yr^{-1}})}=-9.5$. }
    \label{fig:1}
\end{figure}

  We focus on sources at $2<z<4.5$, where the number of sources is sufficient to quantify the average spatial distributions. The relationship between the stellar mass and SFR of SFGs and QGs are shown in Figure \ref{fig:1}. The threshold is located in $\sim1$ dex lower than the main sequence. Also, we see that QGs are mainly selected from the outer envelope of the main sequence of galaxies. It should be noted that this threshold does not select only passive galaxies, which are completely quenched, but post-starburst galaxies as well. The stellar mass of LAEs of SC4K used in this study is reported to have a median value of $\log{(M_\star/M_\odot)=9.0-9.5}$ \citep{Santos2020}, so our photo-$z$-selected galaxies tend to be much more massive than them. 
\par In addition, powerful AGNs, which are detected in X-ray, are not included in this study because the SED fitting may incorrectly estimate their host galaxy properties. We use the X-ray image of Chandra COSMOS Legacy Survey \citep[][]{Civano2016}, which reaches $8.9\times10^{-16}\ {\rm erg\ s^{-1}\ cm^{-2}}$ in $0.5-10$ keV band. We cross-match with the multi-photometric catalog of this survey \citep{Marchesi2016} by using coordinates from optical and NIR images and exclude objects with counterparts with a separation of $\leq1\arcsec$. The fraction of excluded objects is $2-7\%$ and depends the stellar mass and the redshift. Some AGNs that are not bright enough to be detected in this catalog may be included in our sample, but their emission is expected to hardly affect the outcome of the SED fitting.

\subsubsection{Stellar mass completeness}\label{sec:2-1-2}
\par
\begin{figure*}
    \centering
    \includegraphics[width=14cm]{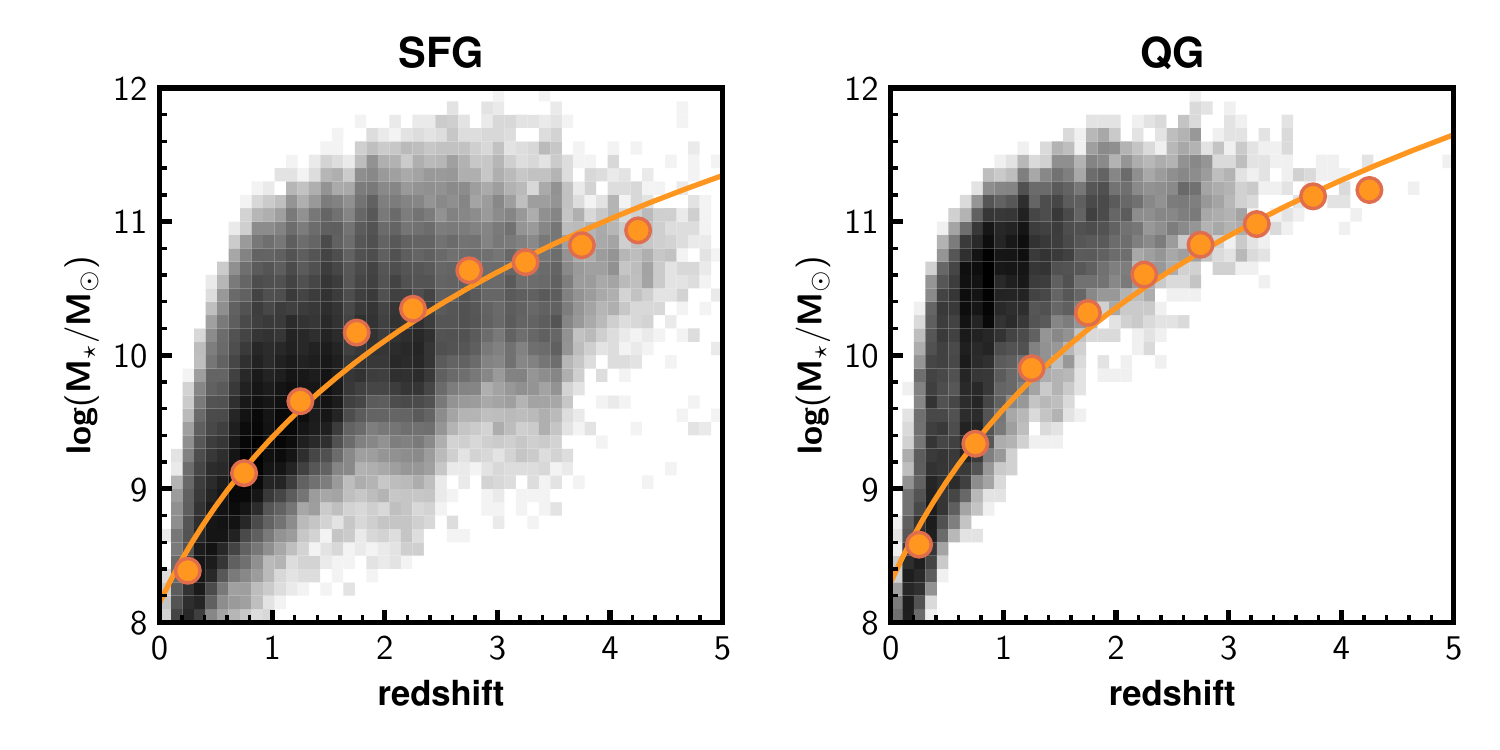}
    \caption{Stellar mass completeness evolution (orange circles) for SFGs (left) and QGs (right). The orange lines show the best power-law fits of observed value in every $\delta z=0.5$. The background color map shows the distribution of observed galaxies with $K_s<24$ mag.}
    \label{fig:2}
\end{figure*}

 We estimate the stellar mass completeness of our SFG and QG sample from the method employed in the previous studies \citep[e.g.,][]{Pozzetti2010,Laigle2016,Davidzon2017}. First, the rescaled stellar mass ($M_{\rm resc}$) expected at the magnitude limit ($K_{s,{\rm lim}}$) is estimated from the $M_\star$ and $K_s$ band magnitude of galaxies. Here, we focus on objects brighter than the limiting magnitude. The $M_{\rm resc}$ is derived as follows:
\begin{equation}
    \log{M_{\rm resc}} = \log{M_\star}+0.4(K_s-K_{s,{\rm lim}}).
\end{equation}
The stellar mass completeness limit is defined as the bottom 90th percentile of the $M_{\rm resc}$ distribution in each redshift bin. We see the evolution of the stellar mass completeness limit of SFGs and QGs in Figure \ref{fig:2}, which shows those in every $\delta z=0.5$. The stellar mass limit is often fitted by the power-law function \citep[e.g.,][]{Davidzon2017}. The best-fits are described as $M_{\star}=10^{8.15}\times(1+z)^{4.11}\ M_\odot$ for SFGs, and $M_{\star}=10^{8.30}\times(1+z)^{4.30}\ M_\odot$ for QGs. The stellar mass completeness limit is generally higher for QGs than for SFGs because the mass-to-light ratios of QGs are higher than those of SFGs. In following sections, we estimate the stellar mass completeness limit for each group following by the above procedure, and employ their value.
\subsection{Ly$\alpha$ emitters}\label{sec:2-2}
\par We use the LAE catalog of \citet{Sobral2018}, who construct a systematic LAE sample at $z\sim2-6$, referred to as the SC4K sample. This catalog is based on the intermediate band (IB) data of Suprime-Cam and narrow band (NB) data of the Wide Field Camera (WFC) of the Isaac Newton Telescope and Suprime-Cam. They select LAEs using an observed equivalent width (EW) threshold and by imposing the color selection and the non-detection of broadband blue-ward Ly$\alpha$ emissions at the target redshift. Spurious objects are excluded from the sample via visual inspection. Contaminants found by previous spectroscopic surveys are also excluded. They finally select 3908 sources as LAEs in 16 redshift slices. 
\par The target redshift range ($2<z<4.5$) corresponds to LAEs selected from NB392, IA427, IA464, IA484, IA505, IA527, IA574, and IA624 bands at $2.20<z<2.24$, $2.42<z<2.59$, $2.72<z<2.90$, $2.89<z<3.08$, $3.07<z<3.26$, $3.23<z<3.43$, $3.63<z<3.85$, and $4.00<z<4.25$, respectively. In this study, we construct a Ly$\alpha$ luminosity ($L_{\rm Ly\alpha}$) complete sample by imposing a $3\sigma$ Ly$\alpha$ luminosity limit cut for each selection band. The $3\sigma$ limiting luminosity of the sample is $\log{(L_{\rm Ly\alpha}/{\rm erg\ s^{-1}})}=42.3-42.8$, dependent on the selection filter \citep[see][for details]{Sobral2018}. We note that NB392 LAEs are selected with a loose threshold (${\rm EW}>5\times(1+z)$ \AA) compared to other IB LAEs (${\rm EW}>50\times(1+z)$ \AA).
\par The spatial coverage of the LAE survey is slightly different from the surveys from which SFGs and QGs are selected. Thus, this study focuses on the region where all of SFGs, QGs, and LAEs exist. We note that the survey fields for most of the selection filters are larger than the field of photo-$z$ galaxies, but that of NB392 is $22\%$ smaller. 
\par It is known that bright LAE can be AGNs \citep[e.g.,][]{Konno2016,Sobral2018}. Such objects can have different properties from those of typical LAEs described in Section \ref{sec:1}. Therefore, LAEs with high Ly$\alpha$ luminosity ($\log{(L_{\rm Ly\alpha}/{\rm erg\ s^{-1}})}>43.3$) are excluded from the sample. The fraction of these objects is quite small ($\sim3\%$ of the total), and if they are included, the following results do not change. We also exclude objects with counterparts in the Chandra COSMOS Legacy Survey catalog in the same manner as that for photo-$z$ galaxies. 
\section{Clustering analysis}\label{sec:3}
\par In this section, we estimate auto- and cross-correlation function signals among SFGs, QGs, and LAEs and discuss the difference in spatial distributions of these galaxy populations.
\subsection{Group construction}\label{sec:3-1}
\par LAEs are constructed in discrete redshifts, as summarized in Section \ref{sec:2-2}. By comparison, the samples of SFGs and QGs have continuous redshift distributions. To increase the signal to noise ratio of cross-correlation functions, we construct four redshift groups at $2<z<4.5$ by combining LAEs selected from eight IBs/NB. These four redshift groups are $2.05<z<2.39$, $2.40<z<2.95$, $2.85<z<3.50$, and $3.50<z<4.50$. Hereafter, we refer to these groups as $z$-group1, $z$-group2, $z$-group3, and $z$-group4, respectively. The redshift range is determined to include possible photo-$z$ galaxies located in the same redshift of LAEs. The redshift range of the lowest redshift bin is defined to match that of NB392 LAEs ($2.20<z<2.24$), including the photo-$z$ uncertainty ($\delta z/(1+z)\sim0.05$), which leads to a narrower range than those of other bins. For the highest redshift group, we select objects within a broader redshift range ($\delta z=1$) to enhance the signal to noise ratio of cross-correlation. Noticeably, a slight duplication exists between the second and the third subgroup, but this does not affect our overall result.
\par Also, we divide SFGs and QGs sample into four subgroups in terms of their stellar mass, which includes only objects whose stellar masses are $\log{(M_\star/M_\odot)}\geq10.4,\ 10.6,\ 10.8,\ {\rm and}\ 11.0$, respectively. Hereafter, we refer to these groups as $M_\star$-group1, $M_\star$-group2, $M_\star$-group3, and $M_\star$-group4, respectively. From the method summarized in Section \ref{sec:2-1-2}, the stellar mass completeness limit of each $z$-group1, 2, 3, and 4 is estimated as $\log{(M_\star/M_\odot)}=10.3,\ 10.6,\ 10.7,\ {\rm and}\ 10.8$ for SFGs and $\log{(M_\star/M_\odot)}=10.6,\ 10.8,\ 11.0,\ {\rm and}\ 11.1$ for QGs, respectively. Some $M_\star$-group whose threshold is below the stellar mass completeness limit are not discussed furthermore. The total numbers of the samples for each of the redshifts and each of the stellar mass thresholds are summarized in Table \ref{tab:1}.
\par
\begin{figure*}
    \centering
    \includegraphics[width=14cm]{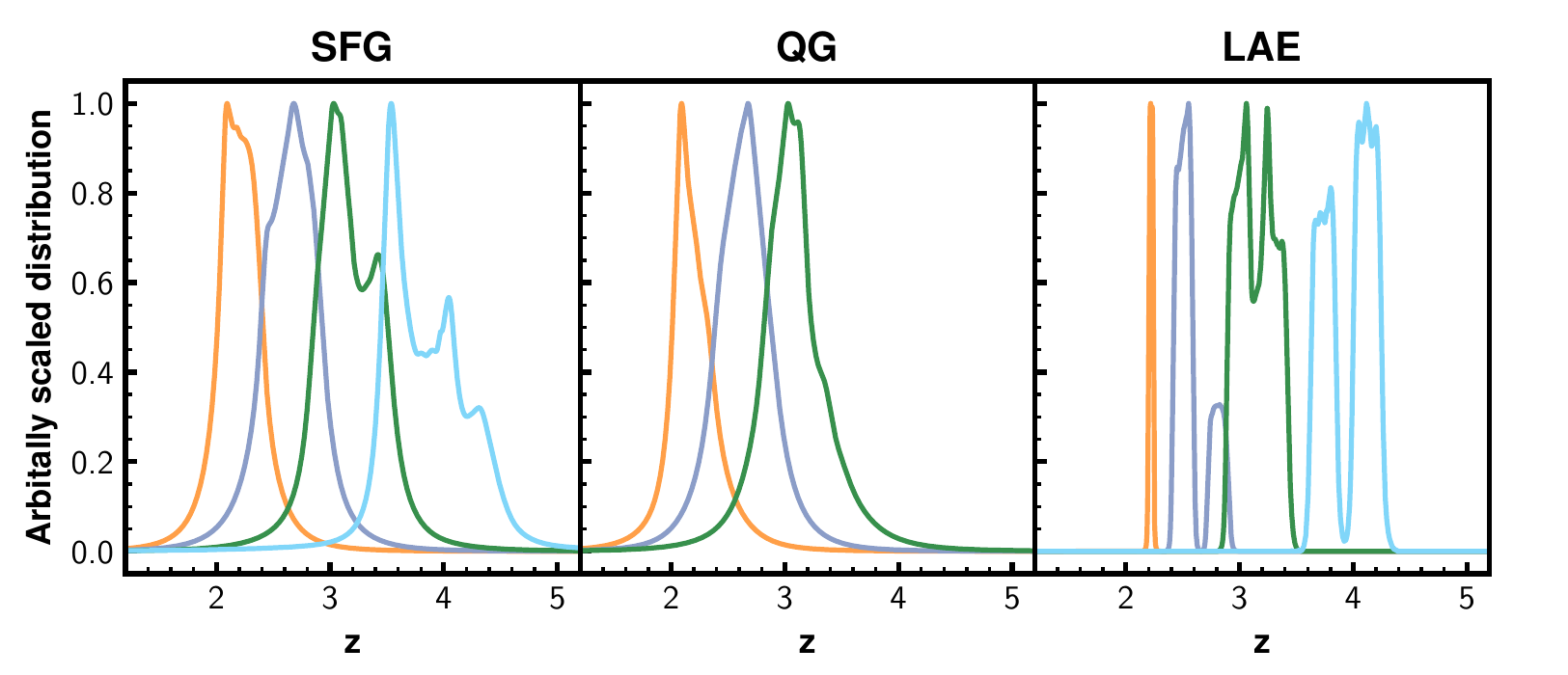}
    \caption{Redshift distributions for each galaxy populations in the four redshift bins. The distributions are normalized at the peak. We show cases of the minimum stellar mass threshold of each redshift groups for photo-$z$ galaxies.}
    \label{fig:3}
\end{figure*}

 The redshift distributions of each group of SFGs and QGs are estimated using the summation of the probability distribution function (PDF) of the photo-$z$ of galaxies in each group, similar to the method used in \citet{Coupon2012}. The photo-z PDF often has a complex shape, but for simplicity, we choose to represent each PDF with normalized Gaussian centered at the median PDF and its 68\% confidence interval as $\pm 1 \sigma$. We sum up these Gaussian function of all objects in a group and construct the average redshift distribution of a group. Meanwhile, the redshift distribution of LAEs is assumed to be number-weighted sum of the Ly$\alpha$ detection rate predicted from each NB/IB transmission curve. The redshift distribution of each sample is summarized in Figure \ref{fig:3}.
\par If the same object exists in two groups wherein the cross-correlation function is measured, the clustering amplitude is artificially increased. Therefore, objects that are also selected as LAEs are excluded from the SFG sample. The fraction of the duplication is $0.05-1\%$ for all groups. There are no QGs classified as LAEs in our sample.

\begin{deluxetable*}{crrrr}
\tabletypesize{\footnotesize}
\tablecaption{Summary of numbers of samples for the clustering analysis. \label{tab:1}}
\tablehead{\colhead{$M_\star$-group}&\colhead{$z$-group1}&\colhead{$z$-group2}&\colhead{$z$-group3}&\colhead{$z$-group4}\\
\colhead{($\log{(M_\star/M_\odot)}$)}& \colhead{$2.05<z<2.39$} &  \colhead{$2.40<z<2.95$} &  \colhead{$2.85<z<3.50$} &  \colhead{$3.50<z<4.50$}}
\startdata
\multicolumn{5}{c}{\textbf{SFG}}\\
$M_\star$-group1 ($>10.4$) & 3077 &    - &     - &    - \\
$M_\star$-group2 ($>10.6$) & 1940 & 2817 &     - &    - \\
$M_\star$-group3 ($>10.8$) & 1128 & 1670 &  1464 &  736 \\
$M_\star$-group4 ($>11.0$) &  561 &  909 &   736 &  371 \\
 \hline
 \multicolumn{5}{c}{\textbf{QG}}\\
$M_\star$-group1 ($>10.4$) &    - &    - &     - &    - \\
$M_\star$-group2 ($>10.6$) &  513 &    - &     - &    - \\
$M_\star$-group3 ($>10.8$) &  421 &  745 &     - &    - \\
$M_\star$-group4 ($>11.0$) &  303 &  587 &   198 &    - \\
 \hline
 \multicolumn{5}{c}{\textbf{LAE}}\\
- &  93 &  725 &  1195 &  161 
\enddata
\end{deluxetable*}
\subsection{Auto-correlation function}\label{sec:3-2}
\par We calculate the auto-correlation function (ACF) using the method from \citet{Landy1993}, who propose an estimator as follows:
\begin{equation}
    \omega_{\mathrm{ACF, obs}}(\theta)=\frac{D D(\theta)-2 D R(\theta)+R R(\theta)}{R R(\theta)},
    \label{eq:1}
\end{equation}
where, $DD,\ DR,$ and $RR$ are the normalized numbers of galaxy--galaxy, galaxy--random, and random--random pairs, respectively.
\par The ACF is often expressed in the power-law form:
\begin{equation}
    \omega_{\mathrm{ACF}} = A_\omega \theta^{1-\gamma}.
    \label{eq:3}
\end{equation}
In this study, we fix $\gamma$ to a fiducial value ($\gamma= 1.8$), following previous studies of LAEs \citep[e.g.,][]{Kusakabe2018, Ouchi2018} and photo-$z$ galaxies \citep[e.g.,][]{Coupon2012}.
\par Random objects are generated with a number density of $\sim 200\ {\rm arcmin^{-2}}$, which is more than 200 times higher than those of photo-$z$ galaxies and LAE. Random objects are distributed in the same region as that of galaxies, including flags.
\par It is known that the observed ACF based on Equation \ref{eq:1} is underestimated because of the finite observation field, referred to as ``integral constraint". To correct this bias, the integral constraint $C$ is derived using the following equation:
\begin{equation}
    C = \frac{\Sigma_i \theta_i^{1-\gamma}RR(\theta_i)}{\Sigma_i RR(\theta_i)}.
\end{equation}
In this study, the $C$ is estimated to be $C=1.46$ for $z$-group1, and $C=1.36$ for the others. Because the survey field is smaller in NB392 (see Section \ref{sec:2-2}), the $C$ is higher in $z$-group1 than in other bins. The $C$ provides a corrected ACF $\omega_{\rm ACF}$ according to the following equation:
\begin{equation}
    \omega_{\rm ACF}= \omega_{\rm ACF,obs}\frac{\theta^{1-\gamma}}{\theta^{1-\gamma}-C}.
\end{equation}
\par The error of the ACFs is estimated based on the Jackknife re-sampling. We divide the observed field into $5\times5$ regions, and removing one region at a time, we estimate $\omega_{i,k}$ in $k$-th trial. This procedure repeats in $25$-times trails and compute the variance of $\omega_{\rm ACF} (\theta_i)$ for each bin:
\begin{equation}
    Var_i= \frac{N-1}{N}\sum_{k=1}^{N} (\omega_{i,k}-\bar{\omega_{i}})^2,
\end{equation}
where $N$ is the number of re-sampling trials, and $\bar{\omega_{i}}$ is the mean of $\omega_{i,k}$. For calculating the variance, we do not consider the trial when the subtracted region is overlapped by over 50\% mask region.
\par We fit a power-law function (Equation \ref{eq:3}) to ACFs via the {\tt Python} module {\tt lmfit} by the least-square method. The distribution of satellite galaxies around the central galaxy is not within the scope of this study, and we are only interested in the larger scale outside of the single halo. Therefore, the angular scale corresponding to within a halo, referred to as the ``one-halo term", are excluded from the ACF fitting range. Specifically, we do not consider $\theta<40\arcsec$ for photo-$z$ galaxies and $\theta<10\arcsec$ for LAEs, where the one-halo term dominates as implied by previous studies \citep[e.g.,][]{Ishikawa2015,Ouchi2018}.
\par We note that the amplitude can decrease due to contaminants. Low-$z$ galaxies whose Balmer breaks can be misclassified as Lyman breaks at the target redshifts in the case of photo-$z$ galaxies. [O{\sc ii}], H$\beta$, [O{\sc iii}] emitters are contaminant candidates for LAEs \citep[see ][for more concrete discussion]{Sobral2018}. Nonetheless, as will be discussed in Section \ref{sec:3-8}, these contaminants do not affect our overall results related to the distribution difference. Therefore, we do not correct these contaminants for the value of the amplitude.
\par
\begin{figure*}
    \centering
    \includegraphics[width=17cm]{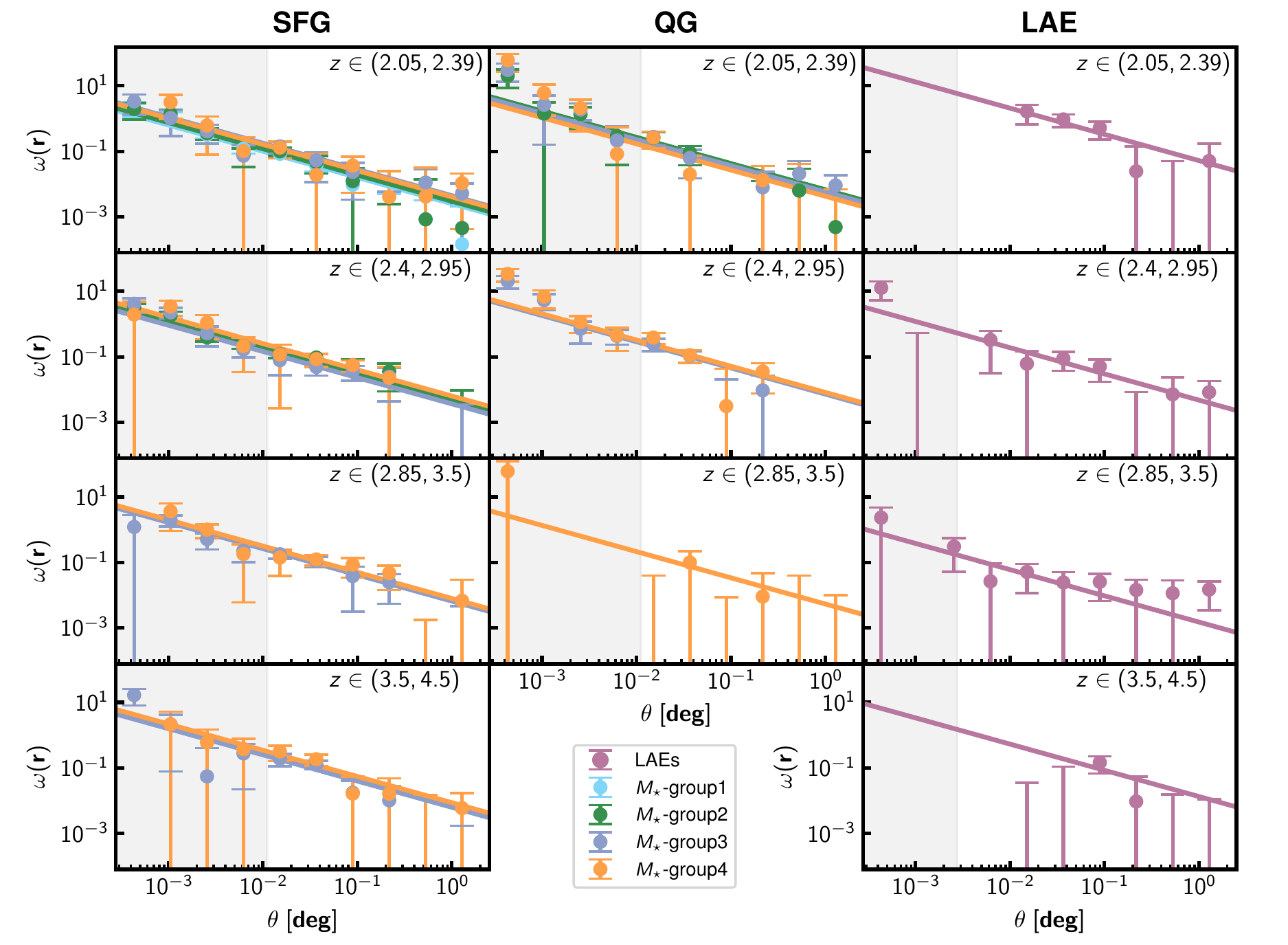}
    \caption{ACFs of SFGs (left), QGs (center), and LAEs (right) at each redshift. Circles show the estimated points, and error bars are derived via Jack-knife re-sampling. Colors among SFGs and QGs represent stellar mass groups. Solid lines correspond to the best fit of identical colors. Gray-shaded regions show ranges not used for deriving the amplitude $A_\omega$ of the power-law.}
    \label{fig:4}
\end{figure*}

 Figure \ref{fig:4} shows the estimated ACFs with the best-fit power-law functions. The observed ACFs are well described in the power-law form in the large scale. Focusing on smaller scale correlation, QGs have a significant deviation from the power-law, especially in $z$-group1 and 2. Such enhancement of the one-halo term for QGs are seen in other studies \citep[e.g.,][]{Cowley2019} and can be related to the higher satellite fraction. The best-fit values of $A_\omega$ are summarized in Table \ref{tab:2}. 
\subsection{Cross-correlation function}\label{sec:3-3}
\par The cross-correlation function (CCF) of samples 1 and 2 is estimated as follows:
\begin{equation}
    \omega_{\mathrm{CCF, obs}}(\theta)=\frac{D_1 D_2(\theta)- D_1 R(\theta)- D_2 R(\theta)+R R(\theta)}{R R(\theta)},
\end{equation}
 where $D_1 D_2,\ D_1 R,$ and $D_2 R$ are the normalized numbers of pair of samples 1 and 2, sample 1 and random, and sample 2 and random, respectively. The integral constraint correction and the error estimation are performed in the same manner as for the ACF.
\par
\begin{figure*}
    \centering
    \includegraphics[width=12cm]{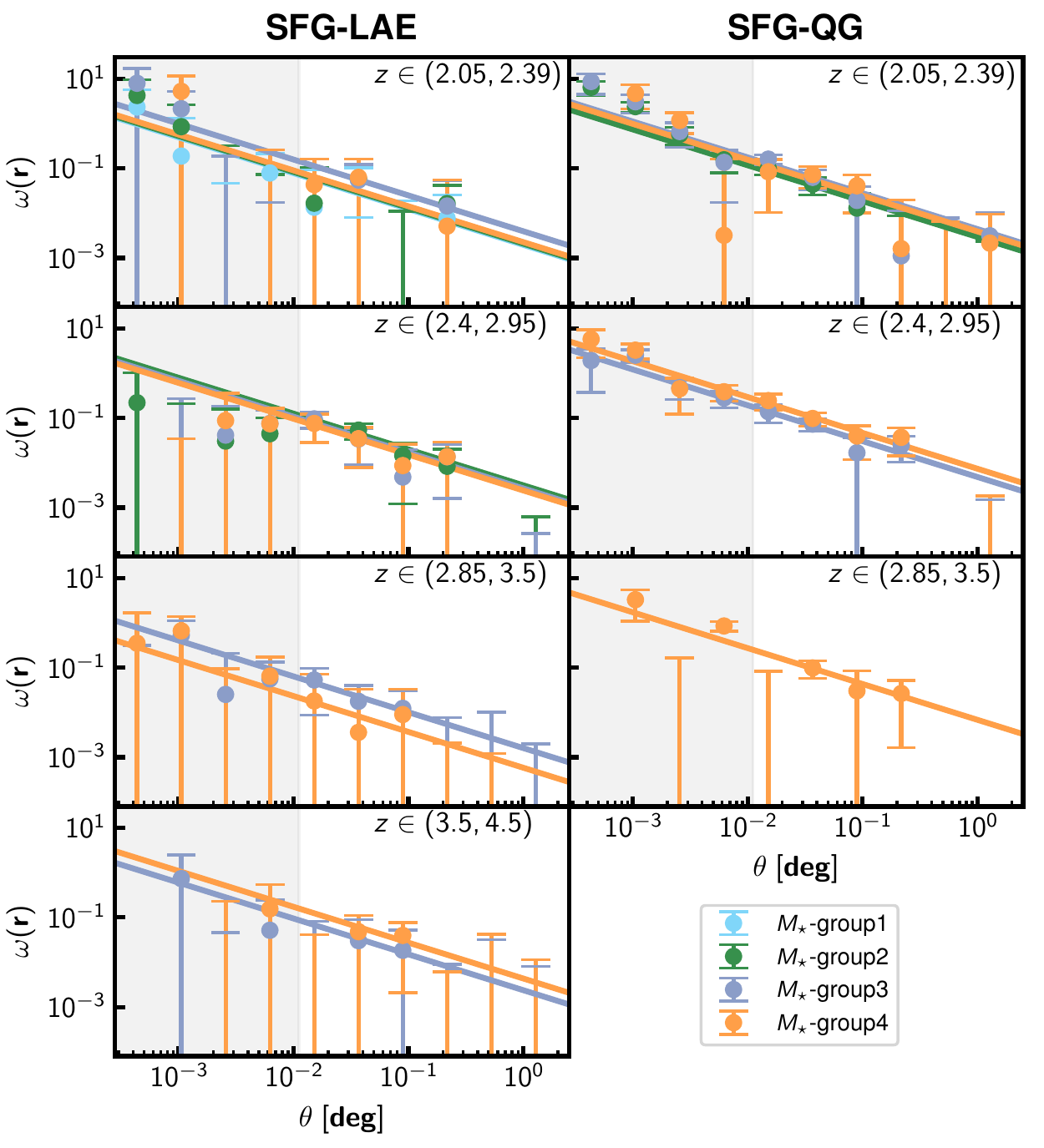}
    \caption{CCFs between SFGs and LAEs (left) and between SFGs and QGs (right) at each redshift. We estimate CCFs between the entire LAE sample and SFGs in each stellar mass bin and those between SFGs and QGs in the same stellar mass bins. Colors show the stellar mass groups of SFGs and QGs, which are identical to those used in Figure \ref{fig:4}.}
    \label{fig:5}
\end{figure*}

 In this study, we measure the difference in the clustering of each population with respect to SFGs. Therefore, we estimate CCFs between SFGs and LAEs and those between SFGs and QGs. The latter cases are determined for the same stellar mass thresholds for both populations. We do not discuss CCFs between QGs and LAEs because we do not obtain any meaningful constraints about the distribution differences due to the poor statistics. The estimated CCFs are shown in Figure \ref{fig:5}. As with ACFs, we fit using the power-law with the fixed $\gamma=1.8$. The CCFs of $\theta>40\arcsec$ are used to avoid the one-halo term for photo-$z$ galaxies. The results are summarized in Table \ref{tab:3}. There is no clear stellar mass dependence of the amplitude of CCF of photo-z galaxies.
\subsection{Correlation length and halo mass}\label{sec:3-4}
\par The ACFs and CCFs are evaluated based on projected separations on the sky. The spatial correlation function $\xi (r)$ can be estimated from ACFs and CCFs. The galaxy spatial correlation function is often approximated as follows:
\begin{equation}
    \xi (r)=\left(\frac{r}{r_0}\right)^{-\gamma},\label{eq:8}
\end{equation}
where $r_0$ is the correlation length. To derive the $r_0$ from ACFs, we employ the Limber equation \citep{1980lssu.book.....P,Efstathiou1991},
\begin{eqnarray}
    A_{\omega}&=&C r_{0}^{\gamma} \frac{\int_{0}^{\infty} F(z) D_{\theta}(z)^{1-\gamma} N(z)^{2} g(z) d z}{[\int_{0}^{\infty} N(z) d z]^{2}}, \label{eq:9}\\
    g(z)&=&\frac{H_0}{c}(1+z)^{2}\left\{1+\Omega_{m} z+\Omega_{\Lambda}\left[(1+z)^{-2}-1\right]\right\}^{1/2},\\
    C&=&\frac{\sqrt{\pi} \Gamma[(\gamma-1)/2]}{\Gamma(\gamma/2)},
\end{eqnarray}
where $D_\theta (z)$ is the angular diameter distance, and $N(z)$ is the redshift distribution of the sample. $F(z)$ describes the redshift evolution of $\xi(z)$, which is modeled as $F(z)=[(1+z)/(1+\bar{z})]^{-3+\epsilon}$ with $\epsilon=-1.2$ \citep{Roche1999}. The $\bar{z}$ is the average redshift of the sample. $c$ and $\Gamma$ are the light speed and the Gamma function, respectively. The derived correlation lengths are summarized in Table \ref{tab:2}.
\par
\begin{figure}
    \centering
    \includegraphics[width=8.5cm]{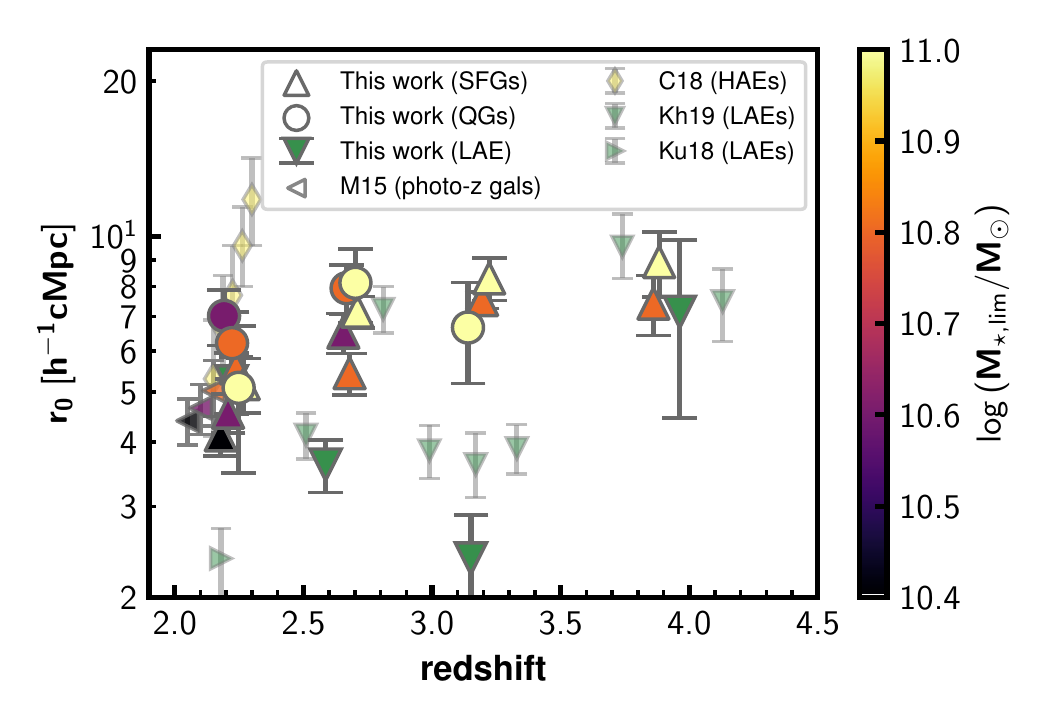}
    \caption{Correlation lengths of different galaxy populations. Colored triangles, circles, and green inverted triangles are correlation lengths of stellar mass groups of QGs, SFGs, and LAEs in this study. Colored left-pointing triangles are those of photo-$z$ galaxies at $2.0<z<2.5$ from \citet{McCracken2015}. Their redshift is slightly shifted for the illustrative purpose. Small green markers are those of LAEs from \citet{Khostovan2019} (inverse triangles) and from \citet{Kusakabe2018} (right-pointing triangle). Small yellow diamonds are those of HAEs with $\log{(M_\star/M_\odot)} >10.1$ \citep{Cochrane2018}. The colors of markers for circles, triangles, and left-pointing triangles show the stellar mass limit of each bins. }
    \label{fig:6}
\end{figure}

 In Figure \ref{fig:6}, the correlation lengths are compared with those in previous studies. We see that our correlation measurement is consistent with previous results within the uncertainty. The correlation length of SFGs at the lowest redshift bin is located in a similar range as that of H$\alpha$ emitters (HAEs) with $\log{(M_\star/M_\odot)}>10.1$ reported in \citet{Cochrane2018}. HAEs are typical star-forming galaxies and expected to have the same correlation length as our SFGs at fixed stellar mass. Those of LAEs at $z\geq2.4$ are also consistent with those in \citet{Khostovan2019}, which also use SC4K LAEs. Though LAEs of $z$-group1 are higher than those at similar redshift in \citet{Kusakabe2018}, their LAEs reach fainter luminosity than ours. Such a difference can cause the different value.  
\par In addition, Figure \ref{fig:6} shows that more massive SFGs have slightly higher amplitudes at fixed redshift, being also consistent with previous studies \citep[e.g.,][]{McCracken2015}. For QGs, we do not see such a trend due to a large uncertainty. 

\par The correlation lengths of the spatial CCF are also derived from the amplitude of its CCFs via the following equation, which is a slightly modified version of Equation \ref{eq:9} \citep{Croom1999}:
\begin{equation}
    A_{\omega}=C r_{0}^{\gamma} \frac{\int_{0}^{\infty} F(z) D_{\theta}(z)^{1-\gamma} N_1(z)N_2(z) g(z) d z}{\int_{0}^{\infty} N_1(z) d z \int_{0}^{\infty} N_2(z) d z}\label{eq:12},
\end{equation}
where $N_1(z)$ and $N_2(z)$ are the redshift distribution functions of samples 1 and 2, respectively. The best-fit correlation lengths are summarized in Table \ref{tab:3}.
\par From the derived correlation length of the spatial ACFs, we also calculate the mean dark-matter halo mass. Firstly, the galaxy-matter bias $b_g$ is estimated as follows:
\begin{equation}
    b_{g}(r)=\sqrt{\frac{\xi(8\ h^{-1}{\rm cMpc})}{\xi_{\mathrm{DM}}(8\ h^{-1}{\rm cMpc}, z)}}, \label{eq:11}
\end{equation}
where $\xi_{\mathrm{DM}}(8\ h^{-1}{\rm Mpc}, z)$ is the correlation function for the dark matter, estimated from the dark-matter power spectrum computed using the transfer function approximation reported in \citet{Eisenstein1998}. Secondly, the mean dark-matter halo mass $\langle M_h \rangle$ is estimated based on an assumption that the mean dark matter halo mass has a galaxy bias equal to the measured value:
\begin{equation}
    b_g = b(\langle M_h \rangle).
\end{equation}
\par In this study, $b(\langle M_h \rangle)$ is based on \citet{Tinker2010}. The uncertainty of $\langle M_h \rangle$ corresponds to a possible range from the $1\sigma$ uncertainty of $b_g$. The derived bias and the mean dark-matter halo mass are also summarized in Table \ref{tab:2}.
\par
\begin{figure}
    \centering
    \includegraphics[width=8.5cm]{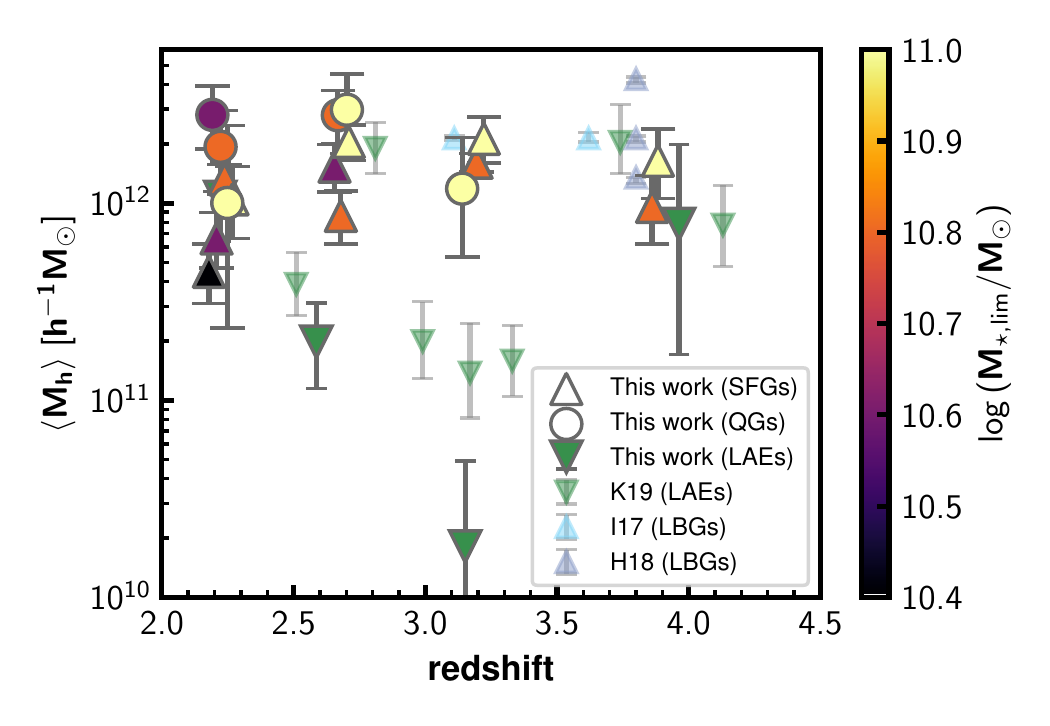}
    \caption{Mean halo mass at $2<z<4.5$. Colored triangles,  colored circles, and green inverted triangles are identical to those in Figure \ref{fig:6}. Small markers are those of Lyman-break galaxies (LBGs) \citep{Ishikawa2017,Harikane2018} and LAEs \citep{Khostovan2019}. }
    \label{fig:7}
\end{figure}

 Figure \ref{fig:7} shows the redshift evolution of the mean dark-matter halo masses of SFGs, QGs, and LAEs. First of all, LAEs tend to reside in less massive halos than SFGs and QGs. Moreover, for SFGs, higher stellar mass galaxies tend to have higher halo masses. This is also inferred from the higher amplitude of ACFs and the findings of previous studies, such as \citet{McCracken2015} at $z\sim2$. For QGs, we do not see such a clear trend due to the large uncertainty compared to those of SFGs, as seen in Table \ref{tab:2}.

\begin{deluxetable*}{cccccc}
\tabletypesize{\footnotesize}
\tablecaption{Summary of the auto-correlation function measurements \label{tab:2}}
\tablehead{ &\colhead{$\log{(M_{\star,{\rm lim}}/M_\odot)}$}\tablenotemark{a}& \colhead{$A_\omega$}\tablenotemark{b} &  \colhead{$r_0$}\tablenotemark{c}&\colhead{$b_g$}\tablenotemark{d}&\colhead{$\langle M_h \rangle$}\tablenotemark{e}\\ &&[${\rm \times10^{-3}deg^{0.8}}$]& [$h^{-1}{\rm cMpc}$]&&[$\times10^{12}h^{-1} M_\odot$]}
\startdata
\multicolumn{6}{c}{SFG}\\
$z$-group1 & 10.4 & $2.53 \pm 0.40$ & $4.12 \pm 0.36$ & $2.01 \pm 0.16$ & $0.45_{-0.14}^{+0.17}$ \\
 & 10.6 & $3.01 \pm 0.47$ & $4.56 \pm 0.40$ & $2.20 \pm 0.17$ & $0.66_{-0.19}^{+0.24}$ \\
 & 10.8 & $4.44 \pm 0.42$ & $5.56 \pm 0.29$ & $2.63 \pm 0.12$ & $1.33_{-0.22}^{+0.24}$ \\
 & 11.0 & $4.07 \pm 0.89$ & $5.18 \pm 0.63$ & $2.47 \pm 0.27$ & $1.05_{-0.38}^{+0.49}$ \\ 
$z$-group2 & 10.6 & $5.07 \pm 0.82$ & $6.52 \pm 0.58$ & $3.45 \pm 0.28$ & $1.53_{-0.39}^{+0.46}$ \\
 & 10.8 & $3.67 \pm 0.61$ & $5.44 \pm 0.50$ & $2.92 \pm 0.24$ & $0.86_{-0.24}^{+0.29}$ \\
 & 11.0 & $6.38 \pm 0.80$ & $7.16 \pm 0.50$ & $3.75 \pm 0.24$ & $2.05_{-0.40}^{+0.45}$ \\
 $z$-group3& 10.8 & $6.58 \pm 0.43$ & $7.54 \pm 0.28$ & $4.46 \pm 0.15$ & $1.61_{-0.17}^{+0.18}$ \\
 & 11.0 & $7.91 \pm 1.35$ & $8.30 \pm 0.78$ & $4.87 \pm 0.41$ & $2.12_{-0.52}^{+0.61}$ \\
$z$-group4 & 10.8 & $6.30 \pm 1.51$ & $7.42 \pm 0.99$ & $5.09 \pm 0.61$ & $0.96_{-0.34}^{+0.43}$ \\
 & 11.0 & $8.69 \pm 2.24$ & $8.92 \pm 1.28$ & $6.01 \pm 0.78$ & $1.64_{-0.59}^{+0.73}$ \\ 
 \hline
 \multicolumn{6}{c}{QG}\\
$z$-group1 & 10.6 & $6.72 \pm 1.48$ & $7.02 \pm 0.86$ & $3.23 \pm 0.36$ & $2.81_{-0.92}^{+1.13}$ \\
 & 10.8 & $5.76 \pm 1.53$ & $6.22 \pm 0.92$ & $2.90 \pm 0.39$ & $1.93_{-0.78}^{+1.02}$ \\
 & 11.0 & $4.26 \pm 2.43$ & $5.09 \pm 1.61$ & $2.42 \pm 0.69$ & $1.00_{-0.77}^{+1.48}$ \\
$z$-group2 & 10.8 & $7.28 \pm 1.44$ & $7.94 \pm 0.87$ & $4.10 \pm 0.40$ & $2.79_{-0.80}^{+0.94}$ \\
 & 11.0 & $8.21 \pm 2.40$ & $8.14 \pm 1.32$ & $4.20 \pm 0.61$ & $2.98_{-1.20}^{+1.54}$ \\
$z$-group3 & 11.0 & $5.37 \pm 2.14$ & $6.67 \pm 1.48$ & $3.92 \pm 0.78$ & $1.19_{-0.65}^{+0.96}$ \\ 
 \hline
 \multicolumn{6}{c}{LAE}\\
$z$-group1 & - & $ 52.18 \pm 12.64 $ & $5.24 \pm 0.71$ & $2.51 \pm  0.30$ & $1.08_{-0.43}^{+0.56}$ \\
$z$-group2 & - & $ 4.79\pm 1.01$ & $3.62 \pm 0.42$ & $1.99 \pm 0.21$ & $0.20_{-0.08}^{+0.11}$ \\
$z$-group3 & - & $ 1.51\pm 0.58$ & $2.38 \pm 0.51$ & $1.58 \pm 0.31$ & $0.02_{-0.01}^{+0.03}$ \\
$z$-group4 & - & $13.41\pm 9.10$ & $7.15 \pm 2.69$ & $5.06 \pm 1.72$ & $0.78_{-0.61}^{+1.20}$ \\
\enddata
\tablenotemark{a}{The stellar mass threshold of the sample.}
\tablenotetext{b}{Best-fit amplitude of power-law (Equation \ref{eq:3}) with $\gamma=1.8$.}
\tablenotetext{c}{Correlation length derived from the Limber equation (Equation \ref{eq:9}).}
\tablenotetext{d}{Galaxy matter bias.}
\tablenotetext{e}{Mean halo mass.}
\end{deluxetable*}
\begin{deluxetable}{cccc}
\tabletypesize{\footnotesize}
\tablecaption{Summary of the cross-correlation function measurements \label{tab:3}}
\tablehead{ &\colhead{$\log{(M_{\star,{\rm lim}}/M_\odot)}$}\tablenotemark{a}& \colhead{$A_\omega$}\tablenotemark{b} &  \colhead{$r_0$}\tablenotemark{c}\\ &&[${\rm \times10^{-3}deg^{0.8}}$]& [$h^{-1}{\rm cMpc}$]}
\startdata
  \multicolumn{4}{c}{SFG-LAE}\\
$z$-group1 & 10.4 & $1.96 \pm 1.12$ & $3.00 \pm 0.95$ \\
 & 10.6 & $2.09 \pm 1.31$ & $3.14 \pm 1.10$ \\
 & 10.8 & $3.95 \pm 0.12$ & $4.40 \pm 0.08$ \\
 & 11.0 & $2.26 \pm 0.91$ & $3.13 \pm 0.70$ \\
$z$-group2 & 10.6 & $3.15 \pm 0.33$ & $4.64 \pm 0.27$ \\
 & 10.8 & $2.69 \pm 0.60$ & $4.23 \pm 0.52$ \\
 & 11.0 & $2.42 \pm 0.37$ & $3.92 \pm 0.34$ \\
$z$-group3 & 10.8 & $1.61 \pm 0.20$ & $3.10 \pm 0.22$ \\
 & 11.0 & $0.59 \pm 0.24$ & $1.77 \pm 0.40$ \\
$z$-group4 & 10.8 & $2.38 \pm 0.26$ & $4.48 \pm 0.27$ \\
 & 11.0 & $4.33 \pm 1.13$ & $6.26 \pm 0.91$ \\
\hline
  \multicolumn{4}{c}{SFG-QG}\\
$z$-group1& 10.6 & $2.94 \pm 0.55$ & $4.50 \pm 0.47$ \\
& 10.8 & $4.42 \pm 0.86$ & $5.50 \pm 0.59$ \\
& 11.0 & $3.91 \pm 0.81$ & $5.00 \pm 0.57$ \\
$z$-group2& 10.8 & $4.84 \pm 0.67$ & $6.34 \pm 0.49$ \\
& 11.0 & $7.39 \pm 0.75$ & $7.73 \pm 0.43$ \\
$z$-group3& 11.0 & $6.88 \pm 0.63$ & $7.83 \pm 0.40$ \\
\enddata
\tablenotemark{a}{The stellar mass threshold of photo-$z$ galaxies.}
\tablenotetext{b}{Best-fit amplitude of power law (Equation \ref{eq:3}) with $\gamma=1.8$.}
\tablenotetext{c}{Correlation length derived from the Limber equation (Equation \ref{eq:12}).}
\end{deluxetable}
\subsection{Distribution differences inferred from correlation functions}\label{sec:3-5}
\par We discuss the difference of distributions of three galaxy populations by comparing spatial correlation functions estimated from ACFs and CCFs. As reported in \citet{Tejos2014}, the following relation among spatial ACFs and CCFs exists according to the Cauchy--Schwarz inequality:
\begin{equation}
    \xi_{D1D2}^2\leq\xi_{D1D1}\xi_{D2D2},
    \label{eq:CS}
\end{equation}
where $\xi_{\rm D1D2}$ is the spatial CCF between samples 1 and 2, and $\xi_{\rm D1D1}$ and $\xi_{\rm D2D2}$ are the spatial ACFs of sample 1 and sample 2, respectively. When the equality is valid, distributions of two populations are determined based only on by their dark matter halo masses, whereas an inequality sign implies the spatial CCF is not determined by halo mass alone and that some additional physics affects their distributions. Therefore, we derive the spatial correlation function ratio $\xi_{\rm D1D2}^2/(\xi_{\rm D1D1}\xi_{\rm D2D2})$ and examine whether or not their distribution is explained only by the dark-matter halo mass. With the assumption of power-law forms of spatial ACFs and CCFs with the same $\gamma$, the ratio is expressed by the correlation lengths:
\begin{equation}
    \frac{\xi_{\rm D1D2}^2}{\xi_{\rm D1D1}\xi_{\rm D2D2}} = \left(\frac{r_{\rm 0,D1D2}^2}{r_{\rm 0,D1D1}r_{\rm 0,D2D2}}\right)^\gamma,
\end{equation}
where $r_{\rm 0,D1D2},\  r_{\rm 0,D1D1},\  {\rm and}\ r_{\rm 0,D2D2}$ are the correlation lengths of the spatial CCF between samples 1 and 2 and their each spatial ACFs, respectively. We use the correlation lengths derived in Section \ref{sec:3-4}.
\par
\begin{figure}
    \centering
    \includegraphics[width=8.5cm]{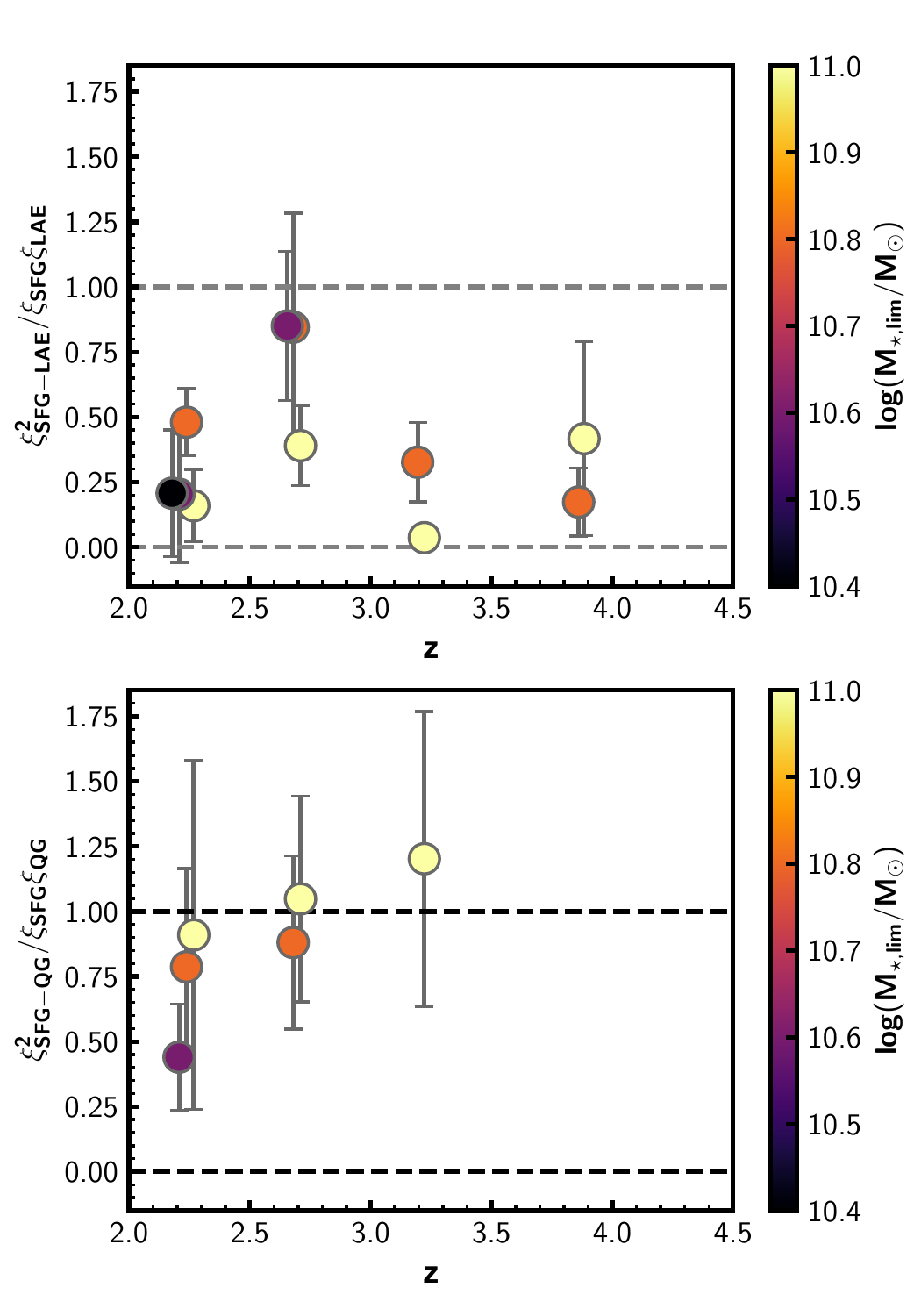}
    \caption{The spatial correlation function ratio $\xi_{\rm D1D2}^2/(\xi_{\rm D1D1}\xi_{\rm D2D2})$. The top panel shows cases for SFGs and LAEs, whereas the bottom panel shows cases for SFGs and QGs. Colors of markers correspond to stellar mass thresholds for SFGs/QGs. The same stellar mass thresholds are imposed for both samples in the case of the correlation function ratio for SFGs and QGs. The redshift is slightly shifted for the illustrative purpose.}
    \label{fig:8}
\end{figure}

 Figure \ref{fig:8} is the main result of this study. The top panel shows the spatial correlation function ratio for SFGs and LAEs. We find that the ratios are below unity for most of the bins, implying that the spatial CCFs between SFGs and LAEs are not determined by the halo mass alone, and that some additional physics segregate the spatial distributions of SFGs and LAEs. This trend is independent of the stellar mass threshold of SFGs. 
\par In the second-lowest redshift ($z\sim2.7$) bin, the spatial correlation function ratio for SFGs and LAEs in $M_\star$-group2 and 3 is consistent with unity if we consider the uncertainty. We do not know the exact origin of the peculiar behavior in this redshift bin, but we show several possibilities. In this redshift bin, LAEs are selected mainly from IA427, leading to a focus on smaller volumes in terms of line of sight compared to those of other bins in higher redshifts. This may induce to trace a peculiar structure by chance. Also, \citet{Cucciati2018} report a ``proto-supercluster" at $z=2.45$ in COSMOS field, within the scope of this bin. This may cause different behavior in that bin. 
\par The limiting magnitude for LAEs is different depending on the selection filter, as mentioned in Section \ref{sec:2-2}. In particular, the limiting magnitude of higher redshift LAEs is shallower. This may cause a bias in the value of each bin. To check this possible bias, we derive the spatial correlation function ratios between SFGs and LAEs brighter than $\log{(L_{\rm Ly\alpha}/{\rm erg\ s^{-1}})}>42.8$ in $z$-group1. This threshold corresponds to the maximum $3\sigma$ limiting luminosity of our LAE sample. The ratios are $0.21\pm0.38$, $0.12\pm0.18$, $0.43\pm0.76$, and $1.54\pm2.56$ for SFGs of $M_\star$-group1, 2, 3, and 4, respectively. Though the last two cases have too large uncertainty to state any trend possibly due to the small sample number of LAEs, these values imply that the limiting Ly$\alpha$ luminosity difference does not impact our results. 
\par The bottom panel of Figure \ref{fig:8} shows the spatial correlation function ratios for SFGs and QGs. Unlike the spatial correlation function ratios for SFGs and LAEs, those for SFGs and QGs maintain unity, suggesting that only their dark-matter halo masses can account for the distributions for SFGs and QGs. If we derive the CCFs of SFGs and QGs adopting the different stellar mass threshold to each population and estimate the ratio, we find that the result generally does not change. 

\subsection{The dependence of the correlation function ratio on the rest-UV magnitude of LAE}
\par To investigate the effect of the stellar mass of LAEs on the results, we divide LAEs in terms of their rest-UV absolute magnitude. The absolute magnitude of SC4K LAEs is calculated based on $i^+$ band photometry summarized in \citet{Laigle2016}, under the assumption of a flat continuum. With the assumption of LAEs locating in the main sequence and with small dust attenuation, the rest-UV luminosity is proportional to SFR and thus to stellar mass. The LAE sample is divided into two subsamples: one with an $M_{\rm UV}$ greater than $-20$, i.e., ``UV-faint LAEs" and the other with an $M_{\rm UV}$ less than $-20$, i.e., ``UV-bright LAEs". LAEs undetected in the $i^+$ band are classified under the former subsample. The 3$\sigma$ limiting magnitude of this $i^+$ band photometry is $26.2$ mag in 3\arcsec aperture \citep{Laigle2016}, which corresponds to $M_{\rm UV}\sim-20.0$ mag at $z=4.5$. This ensures that we completely select UV-bright LAEs at all redshift bins. Moreover, this threshold corresponds to $\log{(M_\star/M_\odot)}\sim9.5-10$ according to the star-formation main sequence and UV magnitude - UV slope relation of SC4K LAEs \citep{Santos2020}. Therefore, in terms of the stellar mass, the UV-bright LAE sample is more similar to QG and SFG than the total LAE sample.
\par
\begin{figure*}
    \centering
    \includegraphics[width=15cm]{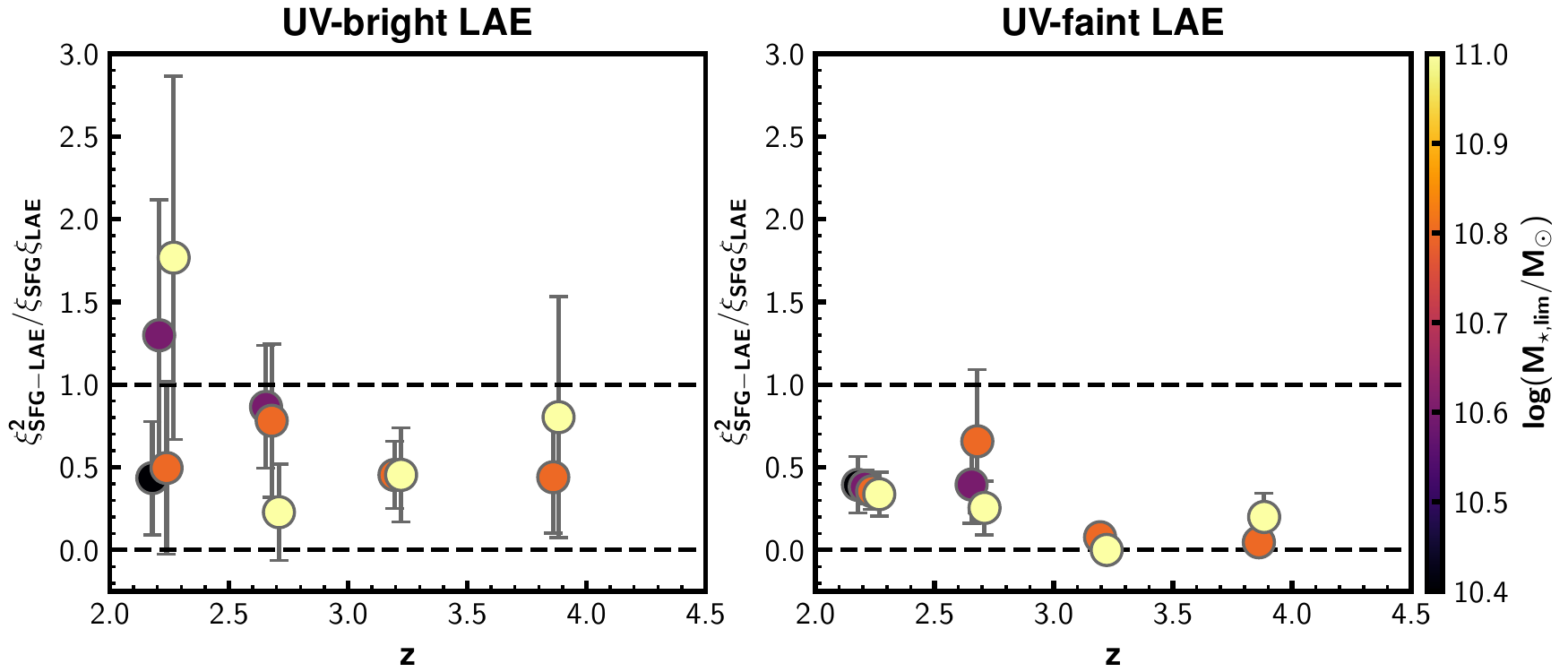}
    \caption{Left panel: The spatial correlation function ratio $\xi_{\rm SFG-LAE}^2/(\xi_{\rm SFG}\xi_{\rm LAE})$ only for UV-bright $(M_{\rm UV}<-20.0)$ LAEs. The marker colors correspond to the stellar mass group of SFGs. Right panel: Same as the left panel, but for UV-faint $(M_{\rm UV}>-20.0)$ LAEs.}
    \label{fig:9}
\end{figure*}

 We derive ACFs and CCFs for these subsamples and estimate the spatial correlation function ratios from the correlation lengths in the same manner as in Section \ref{sec:3-5}. Figure \ref{fig:9} shows the ratio $\xi_{\rm SFG-LAE}^2/(\xi_{\rm SFG}\xi_{\rm LAE})$ as a function of the redshift for the cases of UV-bright LAEs and UV-faint LAEs. Interestingly, the ratios for UV-bright LAEs tend to be higher than those for UV-faint LAEs or have at least same values for some bins within the uncertainty. Moreover, some bins for UV-bright LAEs have ratios equal to unity, suggesting that a distribution difference does not exist, whereas those bins for UV-faint LAEs exhibit ratios less than one. For $z$-group2 cases, this trend may be related to the unity value of the correlation function ratio between the total LAEs and SFGs, but this overall trend implies that the distribution difference between SFGs and LAEs depends on the UV-magnitude of LAEs.

\subsection{Impact of the catastrophic photo-$z$ error}\label{sec:3-8}
\par We have evaluated the impact of the catastrophic error of photo-$z$ on the results. The correlation for $M_\star$-group3 SFGs and LAEs in $z$-group3 bin is used as an example because these have the largest numbers of LAEs, and their Poisson errors do not govern the uncertainty. Galaxies at $0.2<z<0.4$, whose Balmer breaks can be misclassified as Lyman breaks at $z\sim3$, are possible interlopers to our SFG sample at that redshift.
\par
\begin{figure}
    \centering
    \includegraphics[width=8cm]{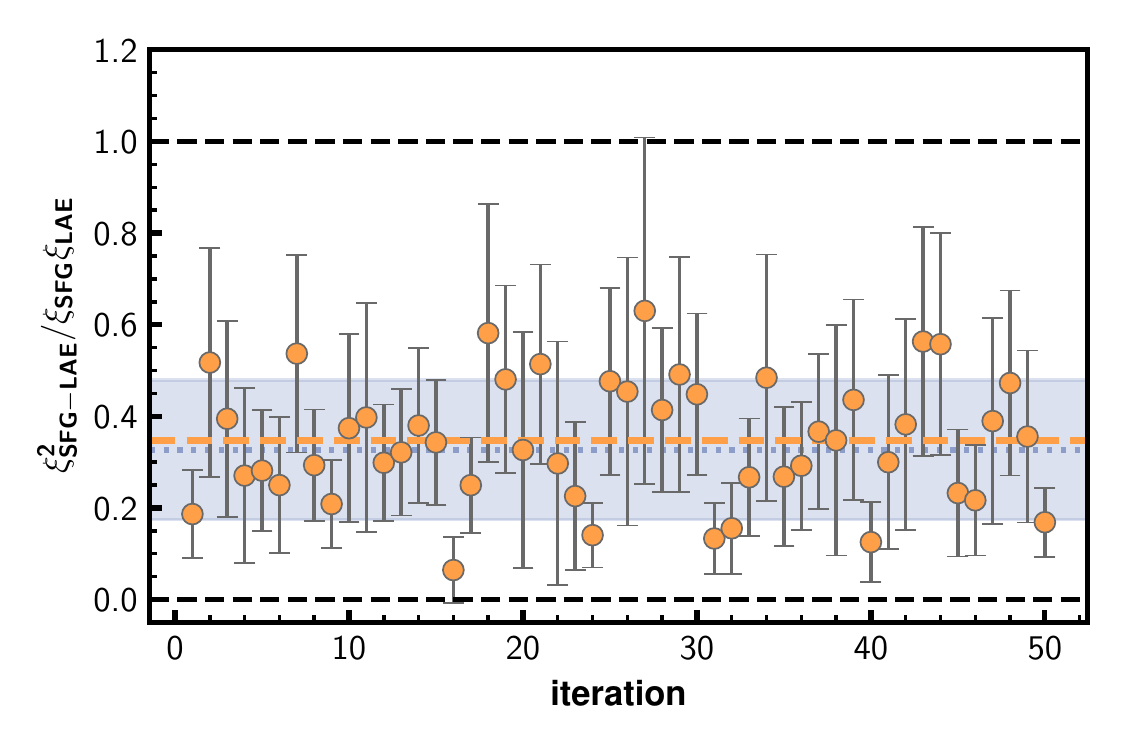}
    \caption{A test of the effect of the catastrophic photo-$z$ error on the trend of the correlation function ratio of SFGs and LAEs. Values of the ratio $\xi^2_{\rm SFG-LAE}/(\xi_{\rm SFG}\xi_{\rm LAE})$ is shown, when 10\% of galaxies are replaced with galaxies at $0.2<z<0.4$, are shown in orange circles. The orange dashed line indicates their average. The blue dotted line and the hatch show the original value and their $1\sigma$ uncertainty range derived in Section \ref{sec:3-5}.}
    \label{fig:10}
\end{figure}

 We randomly select galaxies in our sample and replace them with randomly selected galaxies at $0.2<z<0.4$ from our photo-$z$ catalog. The fraction of the replaced sample corresponds to the contamination fraction of the sample. Although its precise value is not certainly determined, we tentatively assume $10\%$ of our sample. This is of the same order as the fraction of the catastrophic errors of photo-$z$ summarized in \citet{Laigle2016}. We derive ACFs and a CCF between the SFGs and LAEs and a spatial correlation function ratio. We conduct this procedure 50 times in the same manner as in Section \ref{sec:3}. Figure \ref{fig:10} shows the results in terms of the correlation function ratios. The average of the 50-times procedures is consistent with the original value, and the value of individual trials is always below unity. This trend implies that the catastrophic error of photo-$z$ does not change the spatial correlation function ratio. 
\par This trend can be explained by the dependence of the contamination fraction $f$ on the correlation length inferred from the Limber equation (Equation \ref{eq:9} and \ref{eq:12}). The correlation lengths of spatial ACFs decrease by a factor of $(1-f)^{2/\gamma}$, whereas those of spatial CCFs decrease by a factor of $(1-f)^{1/\gamma}$. These factors are compensated in the ratio $\xi^2_{\rm SFG-LAE}/(\xi_{\rm SFG}\xi_{\rm LAE})$ and the ratio equals to the original value. 
\par The same can be applied to the low-$z$ contaminants of LAEs. By matching the spectroscopic redshift in the literature, \citet{Sobral2018} estimate the contamination fraction of SC4K LAEs to be $10-20\%$, which is a similar value to the assumption in the above test. We admit that this is derived from a limited sample; nonetheless, this information supports its insignificant impact on our result.
\section{Overdensity distribution comparison}\label{sec:4}
\par Galaxy overdensity is another often used quantity for characterizing the galaxy environment. \citep[e.g.,][]{Peng2010,Kawinwanichakij2017}. We estimate the overdensity at the positions of SFGs, QGs, and LAEs to examine whether the spatial distribution difference suggested from the clustering analysis can be seen. Because we discuss the spatial distribution difference with reference to SFGs in the clustering analysis, the surface number density of SFGs is used as an index of overdensity.
\par The overdensity at the position of the $i$-th galaxy is defined as follows \citep{Chartab2020}:
\begin{equation}
    \delta (\bm{X}_i) = \sum_{j}\omega_j^i(\frac{\sigma_j(\bm{X}_i)}{\bar{\sigma_j}}-1),
\end{equation}
where $\bm{X}_i$ is the position of the $i$-th galaxy, and $\omega_j^i$ is its probability of residing in the $j$-th redshift slice. $\sigma_j(\bm{X}_i)$ is the surface number density of galaxies at that position, and $\bar{\sigma_j}$ is the average surface number density in the entire field in the $j$-th redshift slice. The redshift slice is generated with an interval of $\delta z/(1+z)=0.01$, which is of the same order as the typical photo-$z$ uncertainty of photo-$z$ galaxies. The $\omega_j^i$ is determined via integration of the PDF distribution of the redshift for the range of each redshift slice. The PDF of photo-$z$ galaxies is assumed to be Gaussian centered at the median PDF and its 68\% confidence interval as $\pm 1 \sigma$. For LAEs, we derive their PDF from the expected Ly$\alpha$ detection rate based on the IB transmission curve.
\par The surface number density map at each redshift slice is estimated using the weighted Gaussian kernel density method. The surface number density at the position of $\bm{X}_i$ is derived as    :
\begin{equation}
    \sigma_{j, {\rm obs}}(\bm{X}_i) = \frac{\sum_{k}\omega_j^k K(\bm{X}_k,\bm{X}_i)}{\sum_{k}\omega_j^k}.
    \label{eq:16}
\end{equation}
This method sums the contributions of all galaxies with the weight $\omega_j^k$ of the probability of $k$-th galaxy being located at $j$-th redshift slice. We consider the 2D Gaussian kernel $ K(\bm{X}_k,\bm{X}_i)$ to be:
\begin{equation}
     K(\bm{X}_k,\bm{X}_i) = \frac{1}{2\pi a^2}\exp{\left[-\frac{r(\bm{X}_k,\bm{X}_i)^2}{2a^2}\right]},
\end{equation}
where $r(\bm{X}_k,\bm{X}_i)$ is the projected distance between two positions, and $a$ is the bandwidth parameter. It is important to carefully select the bandwidth for estimating the adequate scale of the density. Several previous studies determine the bandwidth to minimize the variance of the density map \citep[e.g.,][]{Chartab2020, Badescu2017}, but this leads to bandwidth sizes that differ with redshift. Therefore, we apply a constant bandwidth of 5 cMpc, which is the typical correlation length of SFGs. It is noted that the surface number density at an SFG is systematically higher than the density elsewhere because there is always one galaxy, making it impractical to compare the number density distributions for several populations. Therefore, the contribution from itself is subtracted from $\sigma_{j, {\rm obs}}(\bm{X}_i)$ when the surface number density at an SFG is calculated.
\par Given the finite observed field, it is essential to correct the boundary effect and the masked region. The intrinsic surface number density can be expressed as follows \citep{Jones1993}:
\begin{equation}
    \sigma_{j}(\bm{X}_i) =  \frac{\sigma_{j, {\rm obs}}(\bm{X}_i)}{\int_S K(\bm{X},\bm{X}_i) dS},
    \label{eq:18}
\end{equation}
where $S$ is the area of the observed field with the masking. The denominator in Equation \ref{eq:18} is equal to unity if the position $\bm{X}_i$ is near the center of the observed field and free from the masked region, whereas it becomes smaller if the position $\bm{X}_i$ is at the edge of the field or covered by the mask. We apply the correction for each galaxy.
\par We compare the overdensity distributions of galaxies in $z$-group2 and 3. We focus on this redshift range because it has enough sample numbers for all galaxy populations. We do not consider $z$-group1, because LAEs exist only in a smaller survey field than other samples (see Section \ref{sec:2-2}), which will make it difficult to calculate the density continuously at all redshift range. Furthermore, we assign the same stellar mass threshold to SFGs and QGs, which means that photo-$z$ galaxies with $\log{(M_\star/M_\odot)}>10.8$ in $z$-group2 and those with $\log{(M_\star/M_\odot)}>11.0$ in $z$-group3 are discussed. There is an overlap between the redshift range of these groups, and we conservatively impose the latter threshold for galaxies in that overlapped range. 
\par
\begin{figure}
    \centering
    \includegraphics[width=8.5cm]{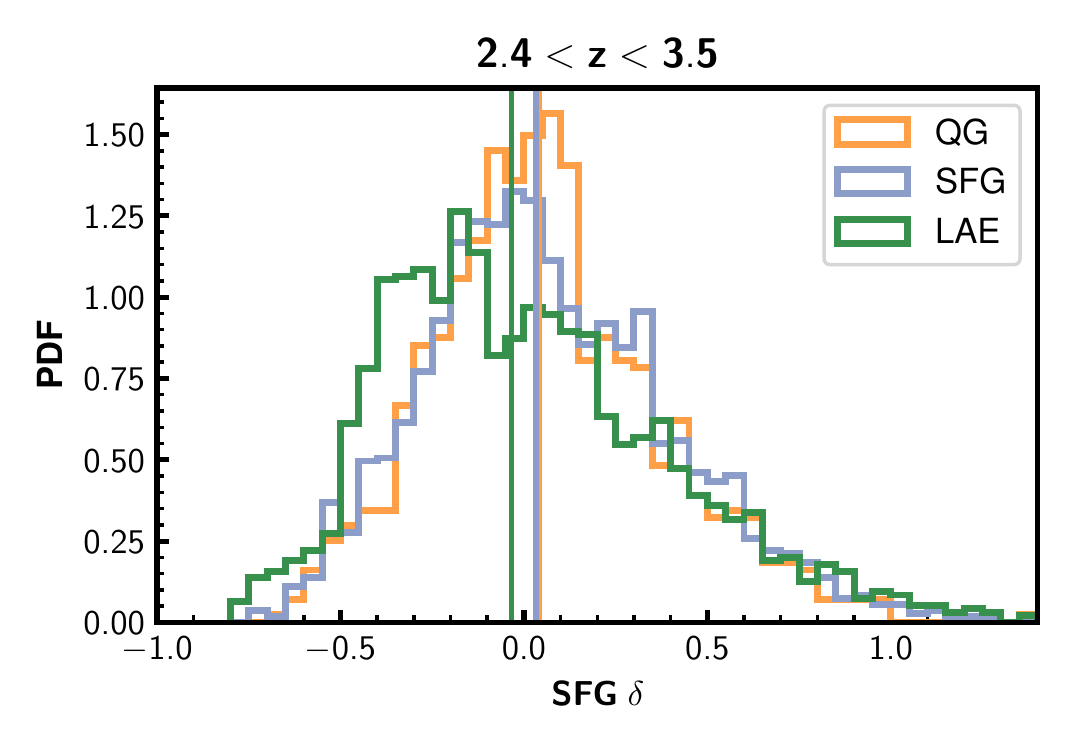}
    \caption{The overdensity distribution of SFGs at the position of QGs (orange), SFGs (blue), and LAEs (green). The median values are shown in vertical lines in colors same as the distribution.}
    \label{fig:11}
\end{figure}

 The overdensity distributions at the position of three galaxy populations are shown in Figure \ref{fig:11}. The overdensity at LAEs tends to be lower than those at SFGs and QGs. We test whether this distribution difference is significant using two statistical tests, the Anderson--Darling (AD) test and the Kolmogorov--Smirnov (KS) test. The AD test is sensitive to the difference at the edge of the distribution, whereas the KS test is sensitive to the difference at the center. The $P$-value from both the AD test and the KS test is less than 0.01, so we reject the null hypothesis that the overdensity distributions of LAEs and SFGs are the same, suggesting that this distribution difference is significant. On the other hand, the overdensity distributions of SFGs and QGs appear to be consistent, suggesting that the QGs are located in a similar environment to SFGs. The statistical tests do not suggest a significant difference between these two overdensity distributions, according to $P=0.10$ from the AD test and $P=0.23$ from the KS test. The median values of the overdensity also support these trends. These suggest that we see the distribution difference between SFGs and LAEs not only from the clustering but also from the overdensity distribution, whereas we do not see it between SFGs and QGs.

\section{Discussion}\label{sec:5}
\subsection{Test of clustering among HAEs, LAEs, and SFGs at $z=2.22$}
\par Thus far, we report that massive SFGs and LAEs are distributed differently beyond the difference of their halo masses. We confirm that the correlation function ratio  $\xi^2_{\rm SFG-LAE}/(\xi_{\rm SFG}\xi_{\rm LAE})$ is less than unity, and this is not affected by the catastrophic failure of the photo-$z$ estimation of some objects. Here, we conduct the same clustering analysis for H$\alpha$ emitters (HAEs) instead of photo-$z$ selected SFGs. HAEs are typical star-forming galaxies more massive than LAEs and have smaller redshift uncertainty than photo-$z$ selected galaxies. Therefore, this test can be used to examine whether or not the trend in Section \ref{sec:3} is caused by the large redshift uncertainty of photo-$z$ selected SFGs. Here, an HAE sample at $z=2.22$ constructed as a part of the HiZELs survey \citep{Sobral2013} is used. This sample was constructed based on the flux excess of ${\rm NB_K}$ at 2.121$\mu{\rm m}$ compared to the K band and the color selection on the $BzK$ diagram. Their redshift uncertainty is as small as that of LAEs, and the survey covers 2.34 ${\rm deg^2}$ in the COSMOS field. Objects with higher fluxes than the average limiting flux $\log{(F_{H\alpha}/ {\rm erg\ s^{-1}\ cm^{-2}})}\sim-16.5$  \citep[Figure 7 in][]{Sobral2013}, at which the completeness of the sample is $\sim50\%$, are used in the study. We use LAEs selected from NB392 from SC4K, which are identical to the sample used in Section \ref{sec:3}, because the selection redshift range is almost the same \citep[Figure 1 in][]{Sobral2017}.
\par The survey area of LAEs is slightly different from that of HAEs. We focus only on regions where LAEs and HAEs coexist and are not affected by any masks of \citet{Laigle2016}. There are three duplications between LAEs and HAEs. For the same reason mentioned in Section \ref{sec:3-1}, we exclude two duplications from HAE sample. The total numbers of HAEs and LAEs are 406 and 87, respectively.
\par
\begin{figure}
    \centering
    \includegraphics[width=8cm]{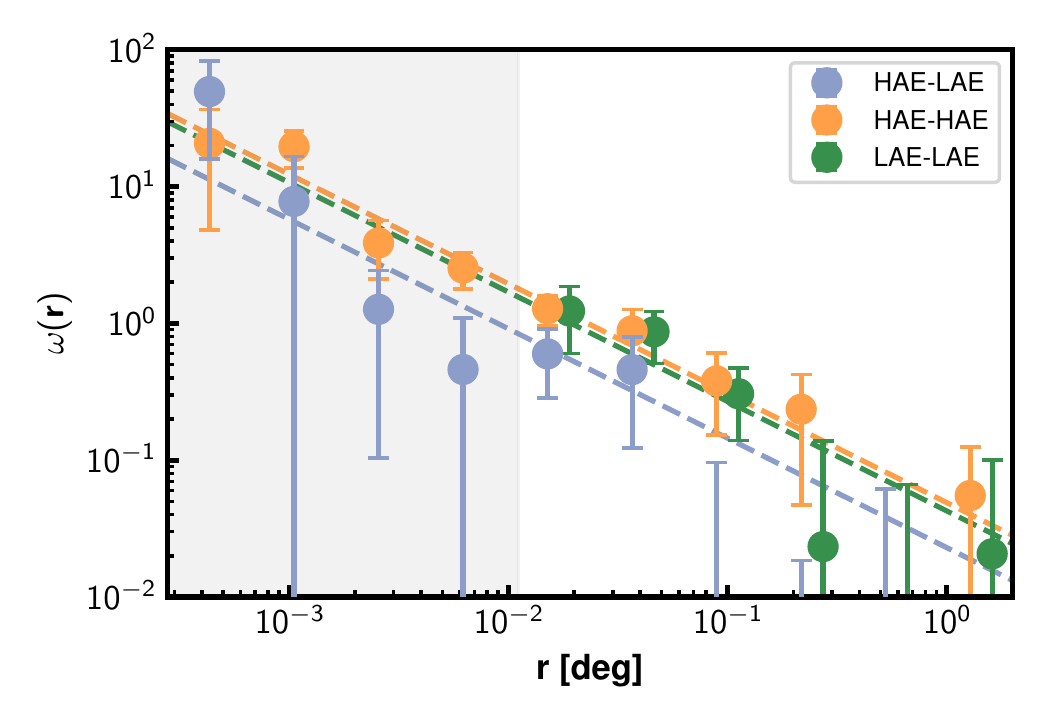}
    \caption{The ACFs and the CCF of HAEs and LAEs at $z=2.22$. Oranges and green circles are ACFs of HAEs and LAEs, respectively, whereas blue circles show their CCF. Best-fit of the power-law (Equation \ref{eq:3}) is shown in dashed lines. The CCF signal is significantly lower than those of ACFs.}
    \label{fig:12}
\end{figure}

 The ACFs and the CCF are estimated in the same manner as in Section \ref{sec:3}. We fit the power-law at $\theta>10\arcsec$ for the ACF of the LAEs and $\theta>40\arcsec$ for the ACF of the HAEs and the CCF. The measured correlation function is shown in Figure \ref{fig:12}. The amplitude of CCF is lower than those of ACFs of HAEs and LAEs. The correlation length of the ACFs and the CCF are estimated in the same manner as in Section \ref{sec:3}, and the spatial correlation function ratio $\xi_{\rm HAE-LAE}^2/(\xi_{\rm HAE}\xi_{\rm LAE})$ is estimated to be $0.27\pm0.12$, which is less than unity. This value suggests that the spatial CCF signal cannot be explained by only their halo mass difference, which is the same as in photo-$z$ SFGs.
\par We also derive the CCFs between HAEs and SFGs. We use SFGs which are in the lowest redshift bin constructed in Section \ref{sec:3} and located in the same survey region as HAEs. In the same way as in Section \ref{sec:3}, four stellar mass thresholds are employed, and ACFs and CCFs are estimated for each. Because both HAEs and SFGs are thought to be similar galaxy populations, many of them are duplicated. Here, we do not exclude these duplications to make the HAE sample analysis consistent with that performed for the LAEs. The values for the ratio $\xi_{\rm HAE-SFG}^2/(\xi_{\rm HAE}\xi_{\rm SFG})$ are calculated to be $0.70\pm0.34$, $0.68\pm0.42$, $0.37\pm0.33$, and $1.35\pm0.84$ for SFGs' stellar mass thresholds of $\log{(M_\star/M_\odot)}>10.4,\ 10.6,\ 10.8,\ {\rm and}\ 11.0$, respectively. These values are higher than the case of HAEs and LAEs, and most of the bins equal unity. This implies that, unlike the clustering between HAEs and LAEs, the clustering between HAEs and SFGs is explainable only by their halo mass.  
\par These tests support the trends shown in Section \ref{sec:3}. Thanks to the smaller redshift uncertainty of HAEs, the spatial correlation function ratio between HAEs and LAEs implies that uncertainties in the photo-$z$ estimates do not cause the trend. The fact that the spatial correlation function ratios for HAEs and SFGs are equal to unity also supports the hypothesis that these differences in distribution do not occur for all line emitters, but only for LAEs. Moreover, these results imply that trends are seen even for less massive SFGs. The H$\alpha$ flux limit corresponds to SFR of $\sim24M_\odot{\rm yr^{-1}}$, based on an assumption of 1 magnitude dust extinction and the standard calibration method by \citet{Kennicutt1998}. Such SFR corresponds to a stellar mass of $\log{(M_\star/M_\odot)}\sim9.8$, according to the relation between SFR and $M_\star$ of HAEs reported in \citet{Oteo2015}, which is $\sim0.6$ dex smaller than the minimum stellar mass threshold of photo-$z$ selected SFGs.
\par It should be noted that the depth of $H\alpha$ flux of this HAE sample has a variation from field to field \citep{Sobral2013,Cochrane2017}. In order to reduce the effect from the depth variance, we verify whether the result with brighter HAEs ($\log{(F_{H\alpha}/ {\rm erg\ s^{-1}\ cm^{-2}})}>-16.0$) is consistent with the original result. The value of the spatial correlation function ratio is consistent within the $1\sigma$ uncertainty both for the correlation function of HAE-LAE and those of HAE-SFG. This suggests that the field variance does not significantly affect our result.

\subsection{Why are LAEs located in a different environment?}\label{sec:5-2}
\par We have investigated the spatial distribution difference among SFGs with $\log{(M_\star/M_\odot)}>10.4$, QGs with $\log{(M_\star/M_\odot)}> 10.6$, and LAEs by two methods, i.e., the clustering analysis and the overdensity analysis. The small signal of CCFs between SFGs and LAEs requires some additional physics to account for it, whereas CCFs between SFGs and QGs can be perfectly explained by their halo mass differences. The CCFs among HAEs, SFGs, and LAEs support the existence of that distribution difference in the case of the stellar mass of SFGs down to $\log{(M_\star/M_\odot)}\sim9.8$ and suggest that the trend is unlikely to be due to the photo-$z$ uncertainty. The overdensity distribution also reveals that LAEs are statistically located in regions that are underdense of SFGs on the scale of $\sim5{\rm cMpc}$, whereas SFGs and QGs are located in the same density field. These trends suggest that LAEs are somehow distributed differently compared to SFGs and QGs.
\par Several previous studies report hints of this distribution difference, indirectly or in peculiar environments. For example, \citet{Momose2020} measure CCFs between H{\sc i} IGM tomography data and several galaxy populations and find that the CCF of LAEs is flat up to $r\sim3h^{-1}$ cMpc, which is different from the behavior of other SFGs. This trend indirectly suggests a potential distribution difference between LAEs and SFGs. In addition, the segregation of LAEs and SFGs is found in protoclusters. \citet{Shi2019} measure the LAE distribution in a known LBG protocluster at $z=3.13$ \citep{Toshikawa2016} and find that LAEs are segregated from the overdensity of LBGs in a few ten cMpc scales. \citet{Shimakawa2017} also report the segregation of LAEs and HAEs in a protocluster core region at $z=2.5$ (please refer to \citet{Hough2020} for perspectives from semi-analytic simulations). This study directly suggests that such distribution segregation between SFGs and LAEs is ubiquitously seen at $z\sim2-4.5$.
\par On the other hand, \citet{Bielby2016} calculate the CCFs between LAEs and spectroscopically confirmed LBGs at $z=3.1$ and demonstrated the ratio $\xi^2_{\rm LBG-LAE}/(\xi_{\rm LBG}\xi_{\rm LAE})=1.28\pm0.46$, which is consistent with unity. The result seems to be inconsistent with ours. However, their spectroscopic confirmation of LBGs is mainly based on Ly$\alpha$ emission or absorption, and the dominant fraction of LBGs seems to have Ly$\alpha$ emission, as seen from their stacked spectra \citep[see Figure 15 in][]{Bielby2012}. This may have lead to tracing similar populations from both samples, which may have caused a higher amplitude in the CCF.
\par The distribution difference between LAEs and SFGs can be explained by the assembly bias \citep[e.g.][]{Gao2006}, which is similar to a scenario suggested in \citet{Shi2019}. LAEs are typically younger in terms of the luminosity weighted age \citep[e.g. approximately 10 Myr in ][]{Nakajima2012,Hagen2014} than massive galaxies, such as the SFGs \citep[e.g. approximately 100 Myr in][]{Hathi2013} or QGs \citep[e.g. approximately 1 Gyr in][]{Belli2019,Gobat2012}. Such differences in age can be related to the different formation time of galaxies and eventually that of their host halos. The different formation epoch of halos is known to cause an impact on the signal of the correlation function. \citet{Zehavi2018} suggest that, even at similar halo mass, the clustering signal can change depending on their formation epoch. The assembly bias tends to increase the clustering signal. If this trend exists in our case, the CCFs between later formed halos and earlier formed halos (i.e., LAEs and SFGs) are expected to be weaker than that expected from each ACFs. We find that UV-brighter LAEs tend to have a higher ratio $\xi^2_{\rm SFG-LAE}/(\xi_{\rm SFG}\xi_{\rm LAE})$ than UV-fainter LAEs. The UV-brighter LAEs are expected to reside in more evolved halos, or in other words, in earlier forming halos, thus reducing such an effect.
\par The large amount of H{\sc i} gas in their circumgalactic media or surrounding intergalactic media associated with massive halos is another possible explanation. This gas absorbs the Ly$\alpha$ photons and prevents us from detecting the Ly$\alpha$ emission of galaxies in massive halos, i.e., galaxies around or in massive halos are preferentially observed as non-LAEs. This makes the distribution of LAEs different from others. Other studies indirectly argue a similar hypothesis. \citet{Toshikawa2016} demonstrate a smaller Ly$\alpha$ equivalent width in an LBG-selected protocluster at $z=3.67$ than in field galaxies. \citet{Shimakawa2017} infer that the accretion of cold streams, which provide pristine H{\sc i} gas to the protocluster core, could prevent Ly$\alpha$ photons from escaping from the dense regions. Meanwhile, \citet{Momose2020} have shown that LAEs tend to avoid H{\sc i} overdensity peaks, whereas \citet{Liang2020} present a similar trend from the correlation of the optical depth of the sightlines of quasi-stellar objects (QSOs) and the spatial distribution of LAEs. We note that these results are on different scales. \citet{Shimakawa2017} show the distribution segregation on a scale of a few hundred pkpc, whereas other studies have focused on a few pMpc. Regardless, the typical scale of this effect and the amount of H{\sc i} gas around massive galaxies remain unclear.
\par It is possible that both effects contribute to the distribution difference. Although a conclusive origin for the distribution difference between SFGs and LAEs remains under debate, our result reinforces the importance of investigating multiple galaxy populations to reveal their environment. 
\subsection{Quenching and environment}
\par
\begin{figure}
    \centering
    \includegraphics[width=8cm]{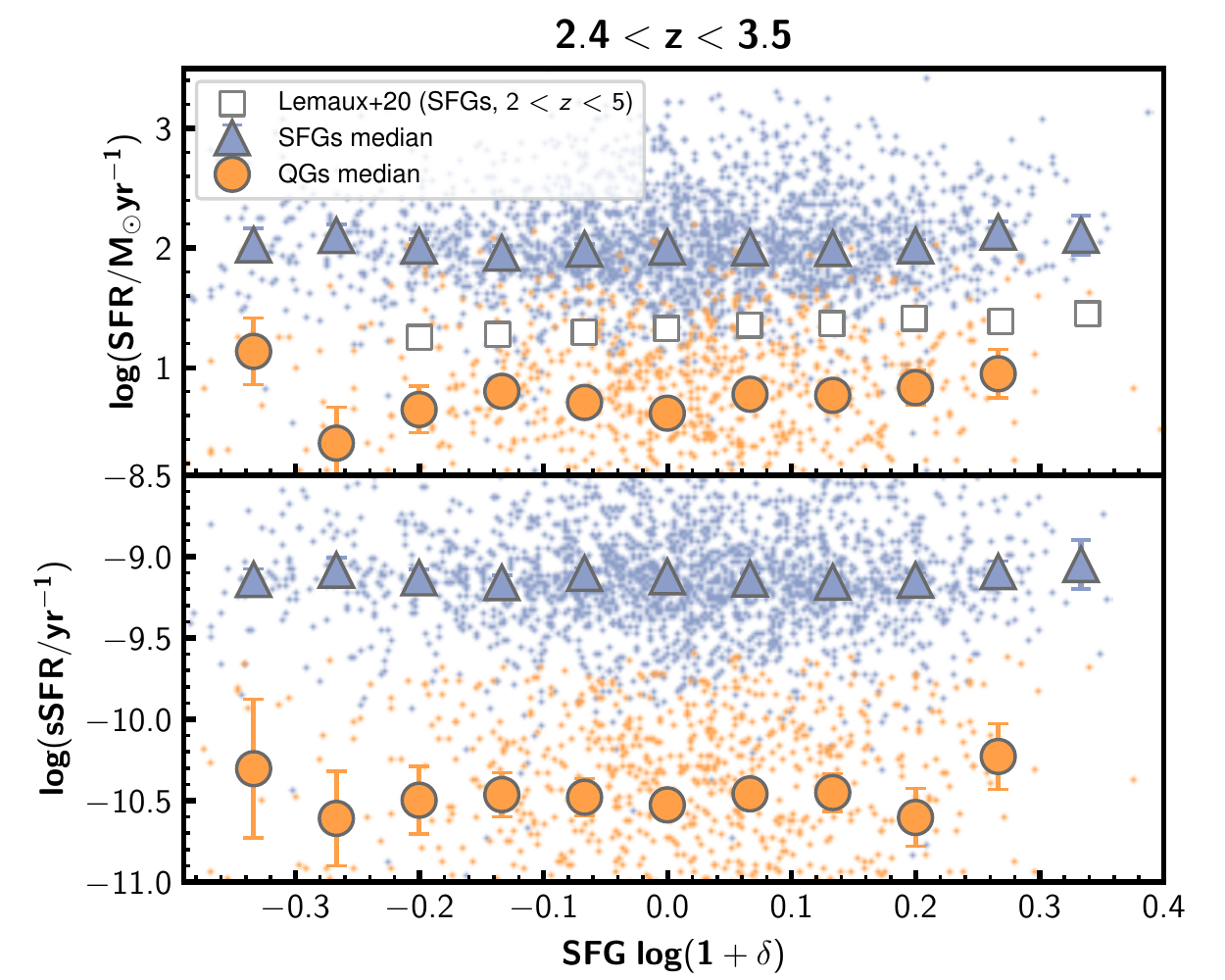}
    \caption{Top panel: Relation of SFR and overdensity for SFGs (blue) and QGs (orange) at $2.4<z<3.5$. Their median values of bins which contain more than ten objects are shown in blue triangles and orange circles. The median trend at $2<z<5$ reported in \citet{Lemaux2020} is shown in white circles for reference. Their median amplitude of SFR is different from ours because of slight difference in their redshift and stellar mass range. Bottom panel: Relation of sSFR and the overdensity. Colors and markers are identical to those of top panel.}
    \label{fig:13}
\end{figure}

 Our sample is large enough to investigate a possible correlation between SFR and the overdensity for different galaxy populations. From the overdensity values for the SFGs and QGs estimated in Section \ref{sec:4}, we check the existence of that correlation at $2.4<z<3.5$. The top panel of Figure \ref{fig:13} shows the relationship between SFR and overdensity. SFGs and QGs are distinguished with their medians and their uncertainties estimated based on the normalized medians of the absolute deviations. The median values may seem to slightly increase towards higher overdensities for SFGs and QGs, especially at $\log{(1+\delta)}>0$, but the Spearman's rank correlation test does not indicate any significant correlations (the correlation coefficient $\rho\sim0.03$ with $P=0.3$ for both populations). Therefore, we conclude that a significant correlation between the number density and SFR is not seen in our sample. Clear trends are also not identified for sSFR, as shown in the bottom panel of Figure \ref{fig:13}. 
\par This result is in contrast to what we observe in the local universe, where there is a clear anti-correlation between SFR and the number density \citep[e.g.,][]{Lewis2002}. Furthermore, at $z>1$, the reversal of the relation has been reported \citep[e.g.,][]{Elbaz2007,Lemaux2020}. In particular, \citet{Lemaux2020} argue the existence of a weak but significant positive correlation between the SFR and the overdensity of star-forming galaxies at $2<z<5$, based on density measurement via Voronoi Monte Carlo mapping. The difference in trends between our results and those of \citet{Lemaux2020}, as shown in Figure \ref{fig:13}, may be due to the different density estimation methods and/or sample difference. Their targets have a lower stellar mass completeness limit (80\% complete to $\log{(M_\star/M_\odot)}\sim9.2-9.5$) than that in our study ($\log{(M_\star/M_\odot)}>10.8$). Moreover, they use a spectroscopically confirmed sample with more accurate redshifts. This can lead to a clearer contrast for the density map and a larger dynamic range for the overdensity. 
\par
\begin{figure}
    \centering
    \includegraphics[width=8cm]{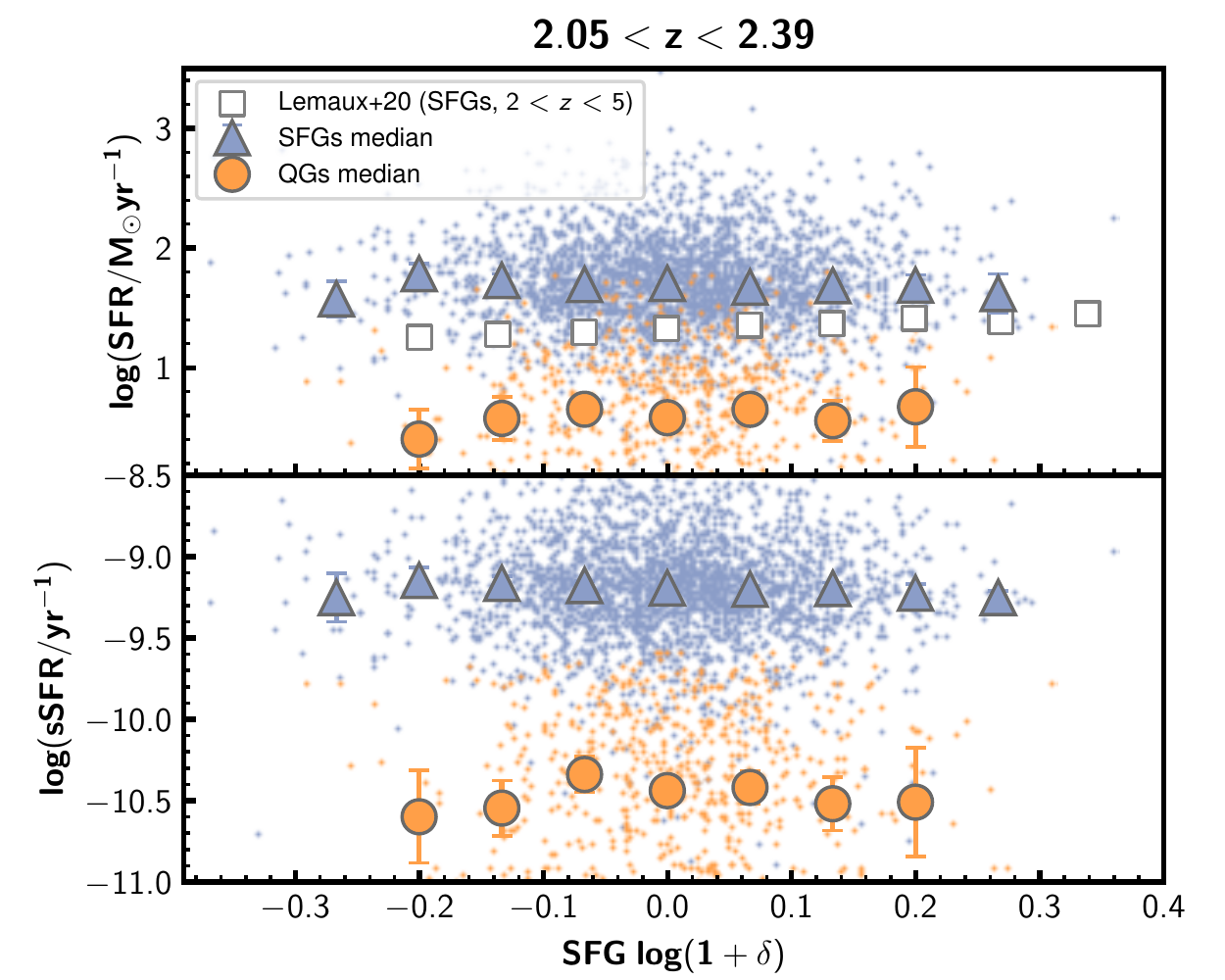}
    \caption{The same as Figure \ref{fig:13}, but for objects in $z$-group1.}
    \label{fig:14}
\end{figure}

 We also check the trend at lower redshift in the same manner and examine the existence of redshift evolution. We use SFGs and QGs in $M_\star$-group2 of $z$-group1 ($2.05<z<2.39$) in Section \ref{sec:3}. Figure \ref{fig:14} shows their SFR-$\delta$ relation. Even a weak increasing trend in median values is not found. The correlation coefficient from Spearman's rank correlation test also does not indicate a significant correlation. The disappearance of the apparently increasing median value trend could be due to the redshift evolution of the relation, which has been reported by \citet{Lemaux2020} at a similar redshift. 
\par Based on the previously presented results, including those from the clustering analysis and the overdensity distribution, our results possibly imply that the environment is not likely to impact significantly on the star-formation quenching of such massive galaxies at $z\geq2$. Several studies support our trends. \citet{Hatfield2017} estimate the CCF signal of SFGs and QGs and argue that the environment does not play a significant role in quenching at $z\sim2$, based on their model for the environmental quenching within the halo occupation distribution scheme. \citet{Lin2016} also report only little dependency on the local density for the quiescent fraction at $1.5<z<2.5$. Also, \citet{Kawinwanichakij2017} argue that mass and environmental quenching are comparable for massive galaxies at $0.5<z<2.0$ with stellar masses similar to those of photo-$z$ galaxies in this study. Their target redshift is lower than ours, so our results may suggest that environmental quenching at $z>2$ is not significant compared to at lower redshift. On the other hand, \citet{Chartab2020} argue that the average SFR of galaxies with stellar masses similar to ours decreases if they are located in more overdense regions, even at $2.2<z<3.5$, suggesting that the galaxy environment does affect quenching at $z>2$. This result may contrast with ours, but this can be related to the different number of the sample or the quality of the SED modeling.
\par Moreover, it should be emphasized that the methods used to measure the environments in previous studies and in this study have large variations in terms of techniques and target scales. For example, \citet{Kawinwanichakij2017} quantify environments based on the 3rd nearest neighbor, which tends to represent a much smaller scale environment than what we explore. \citet{Chartab2020} estimate the number density distribution based on a bandwidth of less than $1h^{-1}$cMpc. Such scale differences may make it difficult to compare the results among different studies.

\section{Summary}
\par In this study, we have investigated the spatial distribution differences among massive ($\log{(M_\star/M_\odot)}\geq10.4$) SFGs and QGs selected by photometric redshift, and LAEs selected by narrow/intermediate bands in the COSMOS field. Through the use of deep and multi-band photometry, a systematic study has been performed for $2<z<4.5$.
\par We first derive the auto-correlation functions and cross-correlation functions of three populations. The spatial correlation function ratio $\xi_{\rm D1D2}^2/(\xi_{\rm D1D1}\xi_{\rm D2D2})$ of SFGs and QGs is equal to unity, suggesting that their distribution can be explained only by their host halo mass. On the other hand, the ratios of SFGs and LAEs are significantly below the unity, implying that some additional physical processes spatially segregate these two populations. This segregation is also implied from the cross-correlation of HAEs and LAEs at $z=2.22$.
\par We also investigate the overdensity at the position of three populations with the use of the surface number density of SFGs. LAEs are found to be located in less dense regions than SFGs and QGs at $2.4<z<3.5$. On the other hand, QGs are confirmed to be located in the same environments as SFGs.
\par With the use of overdensity distribution, we explore the relation between SFR and the overdensity. Neither SFGs nor QGs exhibit significant correlations between SFR and the overdensity. This trend and the results mentioned above suggest that the environment does not significantly impact the star-formation quenching in our dynamic range of the overdensity and the scale of the environment.
\par There are several possible origins of LAEs exhibiting different spatial distributions to other galaxy populations, unlike other galaxy populations. One is assembly bias, which is supported by the higher spatial correlation function ratios of UV-brighter LAEs and SFGs than those of UV-fainter LAEs. The other is a large amount of H{\sc i} gas associated with massive halos in their circumgalactic media or surrounding intergalactic media. Our results highlight the importance of exploring the galaxy environment through multiple populations.
\par Future instruments are expected to help to understand better the results presented in this paper. Firstly, an extensive NIR survey reaching galaxies with less massive stellar masses ($\log{(M_\star/M_\odot)}\sim9$) is essential to reveal the origins of the distribution difference between massive galaxies and LAEs. These breakthroughs should be achievable with future telescopes or instruments such as Nancy Grace Roman Space Telescope or ULTIMATE on the Subaru Telescope. Also, large numbers of spectroscopic samples from the Prime Focus Spectrograph on the Subaru Telescope or the Multi-Object Optical and Near-infrared Spectrograph on the Very Large Telescope will provide us with catalogs of galaxies spanning a wide range of overdensities, leading to a clearer view of the relation between SFR and overdensity and revealing the impact of the assembly bias based on the age derived from their spectrum.
\acknowledgments
We appreciate the anonymous referee for helpful comments and suggestions that improved the manuscript. K.I. acknowledges support from JSPS grant 20J12461. M.K. acknowledges support from JSPS grant 20K14530.

\software{Astropy \citep{Robitaille2013,Price-Whelan2018}, Colossus \citep{Diemer2018}, Matplotlib \citep{Hunter2007}, numpy \citep{Harris2020}, pandas \citep{Mckinney2010}}
\bibliographystyle{aasjournal}

\end{document}